\documentclass[manuscript, screen, 10pt]{acmart}

\usepackage{booktabs} 
\usepackage{amsmath}
\usepackage{amssymb}
\usepackage{listings} 
\usepackage{graphicx}
\usepackage{nameref}
\usepackage{multirow}
\usepackage[inline]{enumitem}
\usepackage{array}
\usepackage{multicol}
\usepackage{graphicx}
\usepackage{diagbox}
\usepackage{xcolor}
\usepackage{listings}
\usepackage{wrapfig}
\usepackage{lscape}
\usepackage{rotating}
\usepackage{epstopdf}
\usepackage{mathtools}
\usepackage{makecell}
\usepackage{natbib}
\usepackage{cleveref}
\usepackage{tikz}
\usepackage[framemethod=TikZ]{mdframed}
\usepackage[bottom]{footmisc}
\usepackage{caption}
\usepackage{subcaption}

\setcitestyle{authoryear,round,citesep={;}}

\usepackage{hyperref}
\hypersetup{
    colorlinks=true,
    linkcolor=blue,
    urlcolor=blue,
    citecolor=blue,
    linktocpage=false
}

\renewcommand\footnotetextcopyrightpermission[1]{}

\lstdefinestyle{scala}
{%
    emph=[1]%
    {%
        groupBy,
        map,
        mapValues,
        union
    },
    emphstyle=[1]{\color{blue}},
    emph=[2]
    {%
        String,
        Iterable,
        Array,
        Set,
        RDD,
        Map,
        Int,
        Long
    },
    emphstyle=[2]{\color{violet}},
}

\lstdefinestyle{scala-types-only}
{%
    emph=[2]
    {%
        String,
        Iterable,
        Array,
        Set,
        RDD,
        Map,
        Int,
        Long
    },
    emphstyle=[2]{\color{violet}},
}

\newcommand{\mynumold}[1]{{\oldstylenums{#1}}}

\newcounter{lstNoteCounter}

\newcommand*\lnnum[1]{\tikz[baseline=(char.base)]{
            \node[shape=circle,draw,inner sep=0.8pt,
                        fill=black, text=white] (char) { \rmfamily\bfseries\footnotesize#1};}}

\lstnewenvironment{annotatedcsource}[3][scala]
{%
    \setcounter{lstNoteCounter}{0}
    \lstset{
        style=#1,
        numbers=right,
        label={#2}, 
        caption={#3},
        escapeinside={(*@}{@*)}}
    }
{}

\begin{document}

\sloppy

\title{Towards Software Analytics: Modeling Maintenance Activities}

\author{Stanislav Levin}
\orcid{0000-0003-4020-7674}
\email{stas.levin@cs.tau.ac.il}
\author{Amiram Yehudai}
\email{amiramy@tau.ac.il}
\affiliation{%
\department{The Blavatnik School of Computer Science}
\institution{Tel Aviv University}
\city{Tel Aviv}
\country{Israel}
}

\begin{abstract}

Lehman's Laws teach us that a software system will become progressively less satisfying to its users over time, unless it is continually adapted to meet new needs. 
Understanding software maintenance can potentially relieve many of the pains currently experienced by practitioners in the industry and assist in reducing uncertainty, improving cost-effectiveness, reliability and more.
The research community classifies software maintenance into 3 main activities: \textit{Corrective}: fault fixing; \textit{Perfective}: system improvements; \textit{Adaptive}: new feature introduction.

In this work we seek to model software maintenance activities and design a commit classification method capable of yielding a high quality classification model.
We performed a comparative analysis of our method and existing techniques based on 11 popular open source projects from which we had manually classified 1151 commits, over 100 commits from each of the studied projects. The model we devised was able to achieve an accuracy of 76\% and Kappa of 63\% (considered ''Good`` in this context) for the test dataset, an improvement of over 20 percentage points, and a relative improvement of $\sim$40\% in the context of cross-project classification.

We then leverage our commit classification method to demonstrate two applications: 
\begin{enumerate*}[label={(\arabic*)}]
    \item a tool aimed at providing an intuitive visualization of software maintenance activities over time, and
    \item an in-depth analysis of the relationship between maintenance activities and unit tests.
\end{enumerate*}

\end{abstract}



\maketitle

\raggedbottom

\section{Software Evolution \& Maintenance}\label{sec:intro}

The software evolution phenomenon was first identified in the late 60's. The term software evolution however, was coined by Lehman only years later \cite{lehman1969programming, lehman1978programs,lehman2003software}. 
Initial studies in this area took place during the 70's and concentrated primarily on measuring and interpreting the growth of software systems and evolutionary trends \cite{lehman2003software, belady1971programming}.
\citet{belady1976model} recognized that the process of large-scale program development and maintenance appeared to be unpredictable, its costs were high and its output was a fragile product. They advocated that one should try to reach beyond understanding and attempt to change the process for the better.
\citet{lehman2000evolution} classify the field of software evolution research into two groups, the first considers the term evolution as a \textit{verb} while the second as a \textit{noun}.
\begin{samepage}
    \begin{description}[font=\normalfont\scshape]
        \item [The verbal view] research is concerned with the question of ``how'', and focuses on means, processes, activities, languages, methods and tools required to effectively and reliably evolve a software system.
        \item [The nounal view] research is concerned with the question of ``what'' and investigates the nature of software evolution, as a phenomenon, and focuses on the nature of evolution, its causes, properties, characteristics, consequences, impact, management and control \cite{lehman2000evolution,lehman2003software}.
    \end{description}
\end{samepage}
\citet{lehman2000evolution,lehman2003software} suggest that both views are mutually supportive. Moreover, they suggest that the verbal view research will benefit from progress made in studying the nounal view, and both are required if the community is to advance in mastering software evolution.

Software maintenance activities are a key aspect of software evolution and have been a subject of research in numerous works \cite{swanson1976dimensions,mockus2000identifying, meyers1988, lientz1978characteristics, levinIcsme2016, schach2003determining}. As a step towards enhanced Software Analytics \cite{buse2010analytics,menzies2013software}, we believe that a better understanding of software maintenance activities could help practitioners reduce uncertainty and improve cost-effectiveness \cite{swanson1976dimensions} by planning ahead and pre-allocating resources towards source code maintenance.
To determine maintenance activity profiles, one must first classify the activities (i.e., developer commits to the version control system), into one of the 3 maintenance activities kinds: \textit{Corrective}: fault fixing; \textit{Perfective}: system improvements; \textit{Adaptive}: new feature introduction.

A widely practiced method for commit classification has been inspecting the commit message\footnote{Also known as ``commit comment''.} \cite{mockus2000identifying, fischer2003populating, sliwerski2005changes, amor2006discriminating}. Works employing commit message based classification reported the accuracy to average below 60\% when used in the scope of a single project, and below 53\% when used in the scope of multiple projects, i.e., when a single model was used to classify commits from multiple projects \cite{hindle2009automatic, amor2006discriminating}. 
Arguably, low accuracy may be a significant barrier preventing these classification methods from being used in professional tools. It would therefore be beneficial to devise maintenance classification methods with higher accuracy (and overall classification quality). 
Our work is also motivated by the following observations:
\begin{enumerate}    
    \item \textit{Cross project classification quality leaves much to be desired.}\\
    Existing results rarely consider cross-project classification, which threatens external validity. \citet{hindle2009automatic} explored cross-project classification and reported the accuracy to be $\sim$52\%, which is considerably lower than the $\sim$60\% range reported by studies dealing with a single project.
    \item \textit{Cohen's Kappa is vital to determine imbalanced classification quality, but it is rarely reported.}\\
    Existing classification results rarely report Cohen's kappa (hence forth Kappa) metric (see also \Cref{sec:statMethods}), which accounts for cases where classification labels (a.k.a classes) are unevenly distributed. Such cases make the accuracy metric somewhat misleading. For example, if the corrective class accounted for 98\% of the commits in a given dataset, and each of the remaining classes accounted for 1\% of the commits, then a simple classification model which always classified commits as corrective would have an impressive accuracy of 98\%. Its Kappa on the other hand, would be 0, making this model much less appealing.
    \item \textit{High quality maintenance activity classification may benefit both previous and future work.}\\
    Our previous work \cite{levinIcsme2016} shows that source code change types as defined by \citet{fluri2006classifying} are statistically significant in the context of maintenance activities defined by \citet{mockus2000identifying}. We believe that increasing the accuracy and Kappa characteristics of commit classification into maintenance activities could improve the quality and accuracy of individual developer maintenance profiles as well as the ability to build predictive models thereof.
\end{enumerate}
In contrast to standard version control systems (VCS) and traditional diff tools which model code changes on the text level, in this work we wish to study changes in object oriented entities such as classes, methods, and fields throughout the life span of a software repository.
To this end we use Fluri's taxonomy of source code changes \cite{fluri2006classifying} for object-oriented programming languages (OOPLs), which consists of 48 (47 $+$ an ''unknown type``) different change types, all of which are project agnostic and describe a meaningful action performed by a developer in a commit (e.g., \textit{statement\_delete}, \textit{statement\_insert}, \textit{removed\_class}, \textit{additional\_class} etc).
Our work explores the following research questions:

\begin{enumerate}[leftmargin=*,labelindent=16pt, label={\textbf{RQ \arabic*.}}]
    
    \item Can fine-grained source code changes be utilized to improve the quality of commit classification into maintenance activities?
    
    \item How does the quality of models which utilize fine-grained source code changes compare to that of traditional models which rely on word frequency analysis only?
    
    \item How can our findings be useful for practitioners and researchers?
\end{enumerate}
This paper is an extension of our previous work \cite{levinPromise2017}, where we first suggested utilizing fine-grained source code changes to classify commits into maintenance activities. 
In this extended paper, we provide a detailed discussion of our commit classification and repository harvesting methods, as well as new perspectives on applications for the discussed methods and techniques.
To that end, \Cref{sec:selectingRepos} provides detailed information about the methods we used to effectively process Big Code, and \Cref{sec:discussion-promise} showcases additional applications which focuses on two particular directions:
\begin{enumerate*}[label={(\arabic*)}]
    \item Software Maintenance Activity Explorer, a tool aimed at providing an intuitive visualization of software maintenance activities over time, and
    \item an in-depth analysis of the relationship between maintenance activities and unit tests in software projects.
\end{enumerate*}

\section{Related Work}\label{sec:relatedWork}

The research community classifies software maintenance into 3 main activities: \textsc{Corrective}, \textsc{Perfective} and \textsc{Adaptive}. The interpretation of these categories, and namely, the criteria to be used to determine which commits fall under what activity type is yet to reach a consensus.
\citet{swanson1976dimensions} and \citet{ghezzi2002fundamentals} suggested the following definitions:
\begin{itemize}
    \item Corrective: rectify the bugs observed while the system is in use.
    \item Perfective: support new features or enhance performance according to user demand.
    \item Adaptive: run on new platforms, new operating systems or interface with new hardware or software.
\end{itemize}

\noindent \citet{mockus2000identifying} used different definitions for the perfective and adaptive activities:
\begin{itemize}
    \item Perfective: code (re-)structuring to accommodate future changes.
    \item Adaptive: new feature introduction.
\end{itemize} 

In this study we adopt the definitions put forth by \citet{mockus2000identifying} and use these definitions to devise a commit classification method that improves existing results.
Having spent almost a decade and a half professionally developing commercial software for both start-ups and enterprises, the authors feel that the definitions suggested by Mockus et al. almost two decades ago, have stood the test of time and remain relevant and applicable to how modern software evolves.
For example, relatively new techniques such as refactoring are now common for improving the quality of code. Despite the fact refactoring became common only years after the definition by Mockus et al. had been suggested, refactoring fits perfectly under their definition for perfective maintenance. The alternative maintenance definitions on the other hand, seem to struggle with accommodating refactoring in a sensible manner.
Moreover, we favour the interpretation by \citeauthor{mockus2000identifying} of the ``adaptive'' maintenance as adding new features (rather than accommodating new operating systems and hardware) since it intuitively covers one of the most basic activities carried out by developers - extending existing software with new features.
The alternative definition of the ``adaptive'' maintenance activity speaks of adapting software to new platforms, operating systems and hardware. We believe that the latter has become significantly less frequent in (modern) software evolution. Even when considering the appearance of smart-phones and other gadgets which required the adaptation of software to new platforms and hardware, the endless stream of new features developers are required to implement in today's software seems like a much more dominant factor.

\citeauthor{mockus2000identifying} suggested the hypothesis that a textual description of the source code change (a commit to the VCS) is essential to understanding why that change was performed. To test this hypotheses, an automatic classification algorithm for maintenance activities was designed based on the textual description of changes. The automatic classification was then verified by surveying 8 developers. The survey results were in line with the automatic classification results, paving the road to text based commit classification approaches. The reported accuracy was $\sim$61\%.
\citet{mockus2000identifying, hindle2009automatic, fischer2003populating, sliwerski2005changes, levinIcsme2016} employed similar, keywords based, techniques for classifying commits into maintenance activities.

Recent work explored using additional information such as commits' author and module, to classify commits both within a single software project, and cross-projects \cite{hindle2009automatic}.
Within a single project, the reported accuracy ranged from $\sim$35\% to 70\% (accuracy fluctuated considerably depending on the project). In a cross-project scope, \citet{hindle2009automatic} reported the classification accuracy to be $\sim$52\%.
A slightly different technique was used by \mbox{\citet{amor2006discriminating}}, who explored classifying maintenance activities in the FreeBSD project by applying a Naive Bayes classifier on commits' comments without an apparent use of keywords. In FreeBSD,  the reported accuracy of classifying a random sample (whose size was not specified) was $\sim$70\% (within the scope of the FreeBSD project).

A summary of the existing results for commit classification into maintenance activities can be found in \Cref{currentResults}.
In this work we were able to improve upon previous results and achieve an accuracy of 76\% and Cohen's kappa of 63\% in the context of cross-project commit classification, an improvement of over 20 percentage points and a relative improvement of $\sim$40\% in accuracy compared to previous results.
 
\begin{table}[ht]    
    \center
    \renewcommand{\arraystretch}{1.5}
    \caption[Classifying commits into maintenance activities, existing results]{Classifying commits into maintenance activities, existing results \cite{hindle2009automatic,amor2006discriminating,mockus2000identifying}}
    \label{currentResults}
    \begin{tabular}{|c|c|c|c|c|}        
    \hline
    \rowcolor{lightgray} \textbf{Study} & \textbf{Scope} &  \textbf{Accuracy} & \textbf{F1 Score} & \textbf{Public Dataset} \\    \hline
   \centering  \multirow{2}{*}{\citet{hindle2009automatic}} & Single Project & 70\% & 0.69 & N/A \\ \cline{2-5}
     & Cross Project &  52\% & 0.51 & N/A \\ \hline
    \citet{amor2006discriminating} &  Single Project & 70\% & N/A & N/A\\ \hline        
    \citet{mockus2000identifying} & Single Project & 61\% & N/A & N/A \\ \hline        
    \end{tabular}    
\end{table}

In contrast to prior studies which typically used the commit message to devise commit classification models, in this work we leverage fine grained source code changes in combination with the commit message to achieve superior model quality.
In addition, we design and evaluate our models in a cross project scope (see also \Cref{tbl:test-dataset-projects}), rather than a single project scope. That is, after performing the per-project stratified sampling to obtain the ground truth dataset (see also \Cref{sec:data-collection}), our subsequent model training and evaluation do not limit the commits to a single project, and are performed on heterogeneous commits (see also \Cref{tbl:test-dataset-projects}).

We also extend our previous work \cite{DBLP:conf/icsm/LevinY17} which studied the co-evolution of test maintenance and code maintenance and showed that maintenance activities can be successfully used to model the number of test methods and test classes in software projects. In particular, we provide statistical evidence showing that software maintenance activities play an important role in modeling test (method and class) counts.

\section{Research Method}

Our research method consists of the following stages:
\begin{enumerate}
    \item Select candidate software repositories and harvest their commit data such as commit message and source code changes performed in the commits (see \Cref{sec:distillingRepos}).
    \item Create a labeled dataset by sampling commits and manually labeling them. Each label is a maintenance activity, i.e. one of the following: corrective, perfective, or adaptive (see \Cref{sec:manualLabels}).
    \begin{enumerate}
        \item Inspect the agreement level on the manually classified commits by having both authors independently classify a 10\% sample of commits (see \Cref{sec:manualAgreement}).
    \end{enumerate}
    \item Devise predictive models that utilize source code changes for the task of commit classification into maintenance activities (see \Cref{modelTypes}).
    \item Evaluate the devised models using two mutually exclusive datasets obtained by splitting the labeled dataset into
    \begin{enumerate*}[label={(\arabic*)}]
        \item a \textit{training} dataset, consisting of 85\% of the labeled dataset, and
        \item a \textit{test} dataset, consisting of the remaining 15\% of the labeled dataset. The test dataset was never used as part of the training process
    \end{enumerate*}
    (see \Cref{sec:evaluaton}).
\end{enumerate}

\subsection{Statistical Methods}
\label{sec:statMethods}
\label{sec:methods}

Picking the optimal classifier for a real-world classification problem is hardly a simple task \cite{fernandez2014we}, however, Random Forest (RF) \cite{ho1998random, breiman2001random} and Gradient Boosting Machine (GBM) \cite{friedman2001greedy,caruana2006empirical,caruana2008empirical} based classifiers are generally considered well performing \cite{caruana2006empirical,fernandez2014we}.
In addition, we also use J48, a variation of the C4.5 \cite{quinlan2014c4} algorithm. The RF implementation \cite{rRF, rRF1} and the GBM's one \cite{rGbm,ridgeway2007generalized} are most likely to outperform the simpler J48 \cite{frank2005weka,rWeka1,rWeka2}, but the latter, in contrast to the formers, is capable of providing a human readable representation of its decision tree. We find this ability valuable since inspecting the decision tree may reveal further insights.
An example of a decision tree produced by the J48 classifier can be found in \Cref{fig:j48-tree}, which depicts our keyword based commit classification model described in \Cref{modelTypes}.

To evaluate the different commit classification models we employ common statistical measures for classification performance. For a given class ${L \in \{\mathit{adaptive},\mathit{corrective},\mathit{perfective}\}}$, 
$\mathit{TP}_L$ is the number of commits correctly classified as class $L$;
$\mathit{FP}_L$ is the number of commits incorrectly classified as class $L$;
$\mathit{FN}_L$ is the number of commits of class $L$ that were incorrectly classified.

\begin{itemize}[itemsep=3pt]
    \item \textit{Precision}$_L$ $ = \frac{\mathit{TP}_L}{\mathit{TP}_L + \mathit{FP}_L}$, the number of commits correctly classified as class $L$, divided by the total number of commits classified as class $L$.
    \item \textit{Recall}$_L$ $ = \frac{\mathit{TP}_L}{\mathit{TP}_L + \mathit{FN}_L}$, the number of commits correctly classified as class $L$, divided by the actual number of $L$ class commits in the dataset. 
    \item \textit{Accuracy} $ = \frac{\sum_{L \in \{a,c,p\}}\mathit{TP}_L}{\sum_{L \in \{a,c,p\}} (\mathit{TP}_L +  \mathit{FP}_L)}$, the proportion of correctly classified commits out of all classified commits.
    \item \textit{No Information Rate (NIR)}, measures the accuracy of a trivial classifier which classifies all commits with using a single class, the one that is most frequent, in our case - corrective.
    \item \textit{Kappa} $ = \frac{\mathit{Accuracy} - \mathit{ExpectedAccuracy}}{1 - \mathit{ExpectedAccuracy}}$, Cohen's kappa, often considered helpful as a measure that can handle  both multi-class and imbalanced class problems (see \Cref{sec:intro}). Cohen's kappa  measures the agreement between the predictions and the actual labels based on both the actual and predicted distributions.
    \item \textit{P-Value [Accuracy $>$ NIR]}, the $p$-value for the null hypothesis that the ''Accuracy $\leq$ NIR`` (i.e., the accuracy of a given predictive model) . A low $p$-value allows one to reject the null hypothesis in favor of the alternative hypothesis that the ''Accuracy $>$ NIR``.
\end{itemize}

\section{Data Collection}

\label{sec:data-collection}
\label{sec:selectingRepos}
\label{sec:distillingRepos}

We use GitHub \cite{gitHub} as the data source for this work due to its popularity \cite{gitHubDominance} and rich query options \cite{gitHubSearch, gitHubSearchNew}.
Candidate repositories were selected according to the following criteria, aimed to capture data-rich repositories that:
\begin{enumerate}
    \item Used the Java programming language (our tools were Java oriented)
    \item Had more than 100 stars (i.e. more than 100 users have ''liked`` these repositories)
    \item Had more than 60 forks (i.e., more than 60 users have ''cloned`` these repositories to their private/organization accounts)
    \item Had their code updated since 2016-01-01 (i.e., these repositories are active)
    \item Were created before 2015-01-01  (i.e., these repositories have existed for several years)
    \item Had size over 2,000 \textsc{KB} (i.e. these repositories are of considerable size)
\end{enumerate}

The criteria aimed at capturing data abundant projects, i.e., projects with plenty of revisions that were still being actively developed.
We found that while popularity related metrics such as stars and forks were a good start, after sampling some of the candidates we identified a number of projects that had little data (revisions) and were therefore not an ideal choice for our study. 
A closer examinations of these projects revealed that more than a few of them turned out to be visually pleasing Android User Interface (UI) controls which had gone viral. To mitigate this, we set a threshold on the repository size in an attempt to filter out small (yet widely popular) projects with little data to analyze.

In light of limited resources we reduced the final candidate set to 11 well known projects from the open source arena, representing various software domains such as IDEs, programming languages (that were implemented in Java), distributed database and storage platforms, and integration frameworks. Following is the list of projected studies in this work (see also \Cref{tbl:prj-info}): 

\begin{enumerate}[topsep=1ex]
    \item \textbf{RxJava} - a library for composing asynchronous and event-based programs for the Java VM.
    \item \textbf{Intellij Community Edition} - a popular IDE for the Java programming language.
    \item \textbf{HBase} -  a distributed, scalable, big data store.
    \item \textbf{Drools} -  a business rules management system solution.
    \item \textbf{Kotlin} - a statically typed programming language for the JVM, Android and the browser by JetBrains.
    \item \textbf{Hadoop} - a framework that allows for the distributed processing of large data sets across clusters of computers.
    \item \textbf{Elasticsearch} - a distributed search and analytics engine.
    \item \textbf{Restlet} - a RESTful web API framework for Java.
    \item \textbf{OrientDB} - a distributed graph database with the flexibility of documents in one product.
    \item \textbf{Camel} - an open source integration framework based on known enterprise integration patterns.
    \item \textbf{Spring Framework} - an application framework and inversion of control container for the Java platform.
\end{enumerate}

\begin{table}[H]
\center
\renewcommand{\arraystretch}{1.5}
\caption[Statistics for the 11 studied projects]{Statistics for the 11 studied projects\footnotemark}
\label{tbl:prj-info}
\begin{tabular}{|l|c|c|}
\hline
\rowcolor{lightgray} \textbf{Project} & \textbf{Total Commits} &  \textbf{Total Contributors} \\\hline
RxJava	& 5,413	& 211 \\ \hline
Restlet &	8,840	& 39 \\ \hline
Drools	& 11,713	& 137 \\ \hline
HBase	& 15,561	& 189 \\ \hline
Spring Framework	& 16,927	& 291 \\ \hline
OrientDb	& 17,035 &	120 \\ \hline
Hadoop	& 19,541	& 137 \\ \hline
Camel	& 32,967	& 410 \\ \hline
Elasticsearch	& 39,958 &	1103 \\ \hline
Kotlin	& 47,386	& 239 \\ \hline
Intellij Community Edition	& 232,607	& 356 \\ \hline
\end{tabular}
\end{table}

\footnotetext{Updated as of 2018, the original study was conducted in 2016.}

Fine-grained source code changes are not directly available in traditional VCSs, Git included, and we therefore had to extract them based on the pre-change and post-change revisions of the changed Java files (which are available in the VCSs).
The task of extracting fine-grained source code changes by comparing two source code files on the abstract syntax tree (AST) level was addressed by the ChangeDistiller \cite{fluri2007change, changeDistillerRepo} and GumTreeDiff \cite{falleri2014fine, gumtree} projects.
Both projects share a common trait, they were designed to operate on two ASTs at a time (typically two subsequent versions of a particular class), and do not support analyzing an entire source code repository's commit history.
In order to distill (harvest) fine-grained source code changes from an entire repository's commit history, our solution design needed to address two main concerns:

\begin{enumerate}[topsep=1ex]
  \item \textbf{Multiple revisions}.
    In the context of modern VCS systems at any given time there is only one revision of each file available in the working tree of a given source code repository. Branches are either a different directory on the file-system\footnote{As implemented in Subversion, see also \url{http://svnbook.red-bean.com/en/1.7/svn.branchmerge.using.html}.}, or require switching to, in which case they swap the current revision for the new one in-place\footnote{As implemented in Git, see also \url{https://git-scm.com/book/en/v2/Getting-Started-Git-Basics}.}.
    Since we are interested in analyzing a given file throughout all its revisions we need to work around this limitation so that for every revision $r$ we have the file's revisions $r$ and $r + 1$ available to the AST comparison tool.
  \item \textbf{Multiple files}.
    A source code repository consists of numerous source code files, created and removed at different points in time throughout the repository's life-cycle. 
    In order to analyze the entire repository an analysis needs to take place for all the source code files (and revisions).
\end{enumerate}

The next stage was to build a mechanism that would replay all the changes made to a given repository according to its commit history so that the fine-grained source code changes could be recorded and repeat this process for every studied repository (see \Cref{lst:distillChangesRepo}). 
The Git VCS system \cite{GIT}, arguably the most popular VCS system in recent years \cite{versionControlPopularity2017, versionControlPopularity2018}, and the one used by the prevalent repository hosting platform \citet{gitHub}, allows one to create a series of patch files, representing the repository's commit history (see also \Cref{lst:prepareRepo}). By applying these patches in a chronological order, one can essentially replay the changes made to a source code repository throughout its commit history (see \Cref{lst:distillChangesPatches,lst:recordPatchContent,lst:distillChangesPair}).

Given that we wish to analyze $n$ repositories, after downloading (cloning) the repositories from GitHub, for each repository $r$ where $1\leq r \leq n$ we created a series of patch files $\{p_{i}^{r}\}_{i=1}^{N_r}$, where $N_r$ is the latest revision number for repository $r$. We only considered the \textit{master} branch\footnote{See also ``Git Branching'', \url{https://git-scm.com/book/en/v1/Git-Branching-What-a-Branch-Is}.}, which is the default branch name in Git. In exceptional cases where the master branch did not exist, we searched for the \textit{trunk} branch, which is the default branch name in Subversion and can sometimes be found in Git repositories that follow Subversion's naming patterns\footnote{See also ``Recommended Repository Layout'', \url{http://svnbook.red-bean.com/en/1.7/svn.tour.importing.html}.}.
Each patch file $p_{i}^{r}$ is responsible for transforming repository $r$ from revision $r_{i-1}$ to revision $r_{i}$, where $r_0$ is the empty repository. By initially setting repository $r$ to revision $1$ (i.e. the initial revision) and then applying all patches ${\{p_{i}^{r}\}}_{i=2}^{N_r}$ in a sequential manner, the revision history for that repository is essentially replayed. Conceptually, this is equivalent to having all developers perform their commits sequentially one by one according to their chronological order.

\vspace{1em}

\noindent
\begin{minipage}{\textwidth}
\begin{lstlisting}[language=Java,label={lst:distillChangesRepo},caption={Distilling fine-grained source changes from multiple repositories}]
distillRepos(repos) {
  for(repo in repos) {
    patches = prepareRepo(repo)
    changes = distillPatches(patches)
    write(changes) // persist the distilled fine-grained changes
  }
}
\end{lstlisting}
\vspace{1em}
\end{minipage}

\noindent
\begin{minipage}{\textwidth}
\begin{lstlisting}[language=Java,label={lst:prepareRepo},caption={Preparing a source code repository for distilling fine-grained source code changes}]
prepareRepo(repo) {
  checkoutRevision(repo, LAST)
  patches = createPatches(repo) // leverage the git-format-patch command
  checkoutRevision(repo, FIRST)
  return orderByPatchId(patches, ASCENDING)
}
\end{lstlisting}
\vspace{1em}
\end{minipage}

\noindent
\begin{minipage}{\textwidth}
\begin{lstlisting}[language=Java,label={lst:distillChangesPatches},caption={Distilling fine-grained source code changes from a sequence of patches}]
distillPatches(patches) {
  for(patch in patches) {
    beforeAfterPairs = recordFileChanges(patch)
    for((revision$_{i}$, revision$_{i+1}$) in beforeAfterPairs) {
        currentChanges = distillChanges(revision$_{i}$, revision$_{i+1}$)
        changes.add(currentChanges)
    }
    return changes
  }
}
\end{lstlisting}
\vspace{1em}
\end{minipage}

\noindent
\begin{minipage}{\textwidth}
\begin{lstlisting}[language=Java,label={lst:recordPatchContent},caption={Recording patch changes}]
recordFileChanges(patch) {
  javaFiles = onlyJavaFilesIn(patch)
  beforeContent = readContent(javaFiles)
  applyPatch(patch) // transform the repo to the next revision
  afterContent = recordContent(javaFiles)
  return zip(beforeContent, afterContent)
}
\end{lstlisting}
\vspace{1em}
\end{minipage}

\noindent
\begin{minipage}{\textwidth}
\begin{lstlisting}[language=Java,label={lst:distillChangesPair},caption={Distilling fine-grained source changes from two files, typically a before-and-after pair}]
distillChanges(left, right) {
  return distillerTool.distill(left, right)
}
\end{lstlisting}
\vspace{1em}
\end{minipage}

We chose ChangeDistiller to perform the fine-grained source code change extraction (i.e., the \verb|distillerTool| in \Cref{lst:distillChangesPair}) due to its popularity in the research community \cite{gall2009change, fluri2008discovering, martinez2013automatically, giger2011comparing, giger2012can, falleri2014fine, fluri2007change, fluri2006classifying, fluri2009analyzing} and its native Java support.
ChangeDistiller required that both the before and after revisions of a source code file were present as physical files on the file system to perform the analysis \cite{changeDistillerAPI}. This design choice presented some challenges in the face of analyzing multiple projects at scale.
Fortunately, ChangeDistiller is an open source tool \cite{changeDistillerRepo} and we were able to easily obtain the source code and surgically resolve this and other issues we encountered.
After the distilling stage was completed, the resulting datasets were manipulated using Apache Spark \cite{sparkSite}, a state of the art framework for large data processing.

Harvesting a real-world software project may yield a great amount of fine-grained source code changes, easily adding up to millions and dozens of millions of records. Manipulating a dataset of this magnitude is no longer as trivial as inputting it into a spreadsheet or even massaging it in a native R environment \cite{R}. As data sizes have outpaced the capabilities of single machines both in terms of memory capacity and CPU speed, users need new frameworks to scale out their computations. As a result, there has been an explosion of new cluster programming models targeting diverse computing workloads \cite{zaharia2016apache} in the ``Big Data'' \cite{diebold2012origin} ecosystem.

Our framework of choice for this work was Apache Spark \cite{sparkSite} (henceforth Spark). Spark has one of the largest developer and user communities\footnote{As indicated by a survey conducted by databricks in 2016, see \url{https://goo.gl/w92BB5}.} and we found its programming model quite intuitive.
It also offers a native Scala language \cite{scala-lang} application programming interface\footnote{Spark provides APIs for a growing number of other programming languages, see \url{https://spark.apache.org/docs/2.3.0/api.html}.} (API), which was a great fit in light of the authors' prior experience with Scala.

One of the fundamental abstractions in Spark is the resilient distributed datasets (RDD) \cite{zaharia2012resilient}. Spark exposes RDDs through a functional programming API where users can pass local functions to run on the cluster (local or distributed). Operations on an RDD are divided into transformations and actions. Transformations derive new RDDs from existing ones, while actions compute and return a concrete result to the program. Spark evaluates RDDs lazily, allowing it to find an efficient plan for the user's computation. In this regard, transformations return a new RDD objects representing the result of a computation but do not immediately compute it. The actual computation takes place when an RDD action is called.

We extensively used Spark to produce \textit{data aggregations} to significantly reduce a dataset's size so it is sufficiently compact to lend itself to interactive exploration in the R environment.
Most of our data aggregations begin with reading all the fine-grained source code changes we have already harvested on a per-project basis and stored them as files on disk, see \Cref{lst:fineGrainedChanges}. 
Transformations are highlighted in blue, Scala type annotations are in violet. Type annotations for local variables can often be omitted in Scala, we explicitly provide them in some of the cases for the sake of clarity.

\begin{minipage}{0.95\linewidth}
\begin{annotatedcsource}{lst:fineGrainedChanges}{Loading all the fine-grained source code changes from the harvested projects}
val sc = new SparkContext(...) // initiate a Spark context

val perProjectData: Set[RDD[Array[String]]] =
    projects
(*@\lnote@*)    .map(prj => sc.textFile(inputNameFor(prj))
(*@\lnote@*)                  .map(line => line.split("#")))

(*@\lnote@*)val fineGrainedChanges: RDD[Array[String]] = sc.union(perProjectData)  
\end{annotatedcsource}
\vspace{1em}
\end{minipage}

The variable \verb|projects| is a collection of project names, over which we iterate and apply a \verb|map| transformation (see bookmark~\lnnum{1}~in~\Cref{lst:fineGrainedChanges}) that builds an RDD from each project's fine-grained source code changes stored as text files on disk. 
Each line in these files is a string concatenation of values separated by a ``$\#$'' (pound) sign. 
We split the lines by the pound sign (see bookmark~\lnnum{2}~in~\Cref{lst:fineGrainedChanges}) so that each element in the resulting RDD is of type \verb|Array[String]|. Since we have multiple projects the \verb|perProjectData| variable is of type \verb|Set[RDD[Array[String]]]|. 
This set of RDDs is then unified into a single RDD for further manipulation using the \verb|union| operation provided by Spark (see bookmark \lnnum{3}). 
Each element in this RDD is an array of strings representing parsed lines from the original files. 
Since RDDs are lazy data structures, no actual processing is done at this point, and it will only take place once an action (e.g., printing, counting, etc.) is invoked on the \verb|fineGrainedChanges| RDD (as indicated in \Cref{lst:fineGrainedChanges-more-grouping-global})

The aggregations we perform on the \verb|fineGrainedChanges| RDD usually fall into one of the following categories:
\begin{itemize}
    \item Per-commit, to explore commit level activity
    \item Per-developer, to explore developer level activity
    \item Per-project, to explore project level activity
    \item Global, to explore the entire dataset's properties
\end{itemize}
For example, to compute the frequencies of the different fine-grained source code changes per commit, i.e., how many times each fine-grained source code change appeared in the commits in our dataset we use the code in \Cref{lst:fineGrainedChanges-by-commit}.

\begin{minipage}{0.95\linewidth}
\begin{annotatedcsource}{lst:fineGrainedChanges-by-commit}{Computing the fine-grained source code change frequencies per commit}
val perCommitFrequencies: RDD[(String, Map[String, Int])] = 
    fineGrainedChanges
(*@\lnote@*)    .groupBy(lineValues => lineValues(COMMIT_ID))
(*@\lnote@*)    .mapValues(countChanges)
\end{annotatedcsource}
\vspace{1em}
\end{minipage}

This computation (\Cref{lst:fineGrainedChanges-by-commit}) uses the \verb|groupBy| and \verb|mapValues| transformations.
The \verb|groupBy| transformation takes an element from the collection it is applied on, i.e., \verb|fineGrainedChanges|, and extracts a key that is used to group all elements with the same key into a single group. Since we would like to compute the frequencies of the different fine-grained source code changes per commit, we first group our records per commit. 
To accomplish this we specify the key to be the commit id\footnote{Also known as ``commit hash'' in git, see also \url{https://git-scm.com/book/en/v2/Git-Basics-Viewing-the-Commit-History}.}. 
This \verb|groupBy| transformation (see bookmark~\lnnum{1}~in~\Cref{lst:fineGrainedChanges-by-commit}) derives a new RDD where each element is a pair of type \verb|(String, Iterable[Array[String]])|. The first tuple component (a.k.a ``key'') is the commit id, and the second (a.k.a ``value'') is a collection of all the elements that had this particular key. Next we apply a \verb|mapValues| transformation (see bookmark~\lnnum{2}~in~\Cref{lst:fineGrainedChanges-by-commit}) which iterates over these pairs and transforms their value while retaining the key. The transformation logic we provide to \verb|mapValues| is one that calculates the frequencies of each fine-grained source code change, see \Cref{lst:count-frequency}.

\begin{minipage}{0.95\linewidth}
\begin{annotatedcsource}[scala-types-only]{lst:count-frequency}{Computing the fine-grained source code change frequencies}
def countChanges(lines: Iterable[Array[String]]): Map[String, Int] =
    lines
(*@\lnote@*)    .map(lineValues => lineValues(CHANGE_TYPE))
(*@\lnote@*)    .groupBy(identity)
(*@\lnote@*)    .mapValues(_.size)
\end{annotatedcsource}
    \vspace{1em}
\end{minipage}

The \verb|countFrequencies| is a method which receives an iterable of lines and represents all changes performed in a given commit, it returns a mapping (\verb|Map[String, Int]|) between the fine-grained source code change type (e.g., {\footnotesize ``ADDITIONAL\_CLASS''}) and its frequency. 
Note that \verb|countFrequencies| does not operate on RDDs but on Scala native collections. One of the benefits of using Spark's Scala API is that it is consistent with Scala's native collections. In particular, the name and semantics of the \verb|mapValues| and \verb|groupBy| transformations for Scala collections and Spark RDDs are the same.

First each line is mapped to its corresponding fine-grained source code change type (bookmark~\lnnum{1}~in~\Cref{lst:count-frequency}), then all values are grouped using the \verb|identity| key extractor (bookmark~\lnnum{2}~in~\Cref{lst:count-frequency}), forming tuples where the key is the fine-grained source code type and the value is a collection of all the corresponding fine-grained source code change types equal to the key. Finally, we map the tuples' values (bookmark~\lnnum{3}~in~\Cref{lst:count-frequency}) to the sizes of their value component. 
This results in tuples where the key is the fine-grained source code change type, and the value is the key's frequency. 
Since the keys in these tuples are the fine-grained source code change types, we end up with a mapping (\verb|Map[String, Int]|) between the fine-grained source code change type (e.g., {\footnotesize ``ADDITIONAL\_CLASS''}) and its frequency.

\noindent 
For example, if a project's raw data file contains the following pound separated values:

\begin{lstlisting}[mathescape=true,numbers=none,frame=none,basicstyle=\ttfamily\footnotesize,xleftmargin=0em,backgroundcolor=\color{white}]
1a2b3c<@{\Large\textbf{\#}@>PARAMETER_INSERT<@{\Large\textbf{\#}@>file1.java
1a2b3c<@{\Large\textbf{\#}@>ADDITIONAL_FUNCTIONALITY<@{\Large\textbf{\#}@>file3.java
1a2b3c<@{\Large\textbf{\#}@>DOC_DELETE<@{\Large\textbf{\#}@>file2.java
1a2b3c<@{\Large\textbf{\#}@>PARAMETER_INSERT<@{\Large\textbf{\#}@>file1.java 
1a2b3c<@{\Large\textbf{\#}@>PARAMETER_INSERT<@{\Large\textbf{\#}@>file1.java 
1a2b3c<@{\Large\textbf{\#}@>DOC_DELETE<@{\Large\textbf{\#}@>file2.java
\end{lstlisting}

The \verb|map| transformation (bookmark~\lnnum{1}~in~\Cref{lst:count-frequency}) results in:

\begin{lstlisting}[mathescape=true,numbers=none,frame=none,basicstyle=\ttfamily\footnotesize,xleftmargin=0em,backgroundcolor=\color{white}]
{PARAMETER_INSERT}
{ADDITIONAL_FUNCTIONALITY}, 
{DOC_DELETE}
{PARAMETER_INSERT}
{PARAMETER_INSERT}
{DOC_DELETE}
\end{lstlisting}

The \verb|groupBy| transformation (bookmark~\lnnum{2}~in~\Cref{lst:count-frequency}) results in:

\begin{lstlisting}[mathescape=true,numbers=none,frame=none,basicstyle=\ttfamily\footnotesize,xleftmargin=0em,backgroundcolor=\color{white},xleftmargin=0em,xrightmargin=0em,framexleftmargin=0em,framexrightmargin=0em]
(PARAMETER_INSERT         -> {PARAMETER_INSERT, PARAMETER_INSERT, PARAMETER_INSERT})
(ADDITIONAL_FUNCTIONALITY -> {ADDITIONAL_FUNCTIONALITY})
(DOC_DELETE               -> {DOC_DELETE, DOC_DELETE})
\end{lstlisting}

The \verb|mapValues| transformation (bookmark~\lnnum{3}~in~\Cref{lst:count-frequency}) results in:

\begin{lstlisting}[mathescape=true,numbers=none,frame=none,basicstyle=\ttfamily\footnotesize,xleftmargin=0em,backgroundcolor=\color{white},xleftmargin=0em,xrightmargin=0em,framexleftmargin=0em,framexrightmargin=0em]
(PARAMETER_INSERT         -> 3)
(ADDITIONAL_FUNCTIONALITY -> 1)
(DOC_DELETE               -> 2)
\end{lstlisting}

The {\ttfamily{perCommitFrequencies}} RDD (see \Cref{lst:fineGrainedChanges-by-commit}) will therefore contain the element:

\begin{lstlisting}[mathescape=true,numbers=none,frame=none,basicstyle=\ttfamily\footnotesize,xleftmargin=0em,backgroundcolor=\color{white},xleftmargin=0em,xrightmargin=0em,framexleftmargin=0em,framexrightmargin=0em]
(1a2b3c -> {PARAMETER_INSERT -> 3, ADDITIONAL_FUNCTIONALITY -> 1, DOC_DELETE -> 2})
\end{lstlisting}

Per-developer and per-project aggregations are performed similarly to what we have shown for the per-commit aggregation, the main change being the key passed to the \verb|groupBy| transformation (see~bookmarks~\lnnum{1}~and~\lnnum{2}~in~\Cref{lst:fineGrainedChanges-more-grouping}).
Global operations require no prior aggregations and can be performed directly on the \verb|fineGrainedChanges| RDD, see \Cref{lst:fineGrainedChanges-more-grouping-global}.

\begin{minipage}{0.95\linewidth}
\begin{annotatedcsource}{lst:fineGrainedChanges-more-grouping}{Per-developer and per-project aggregations}
// aggregate per developer (email) and project
val perDeveloper: RDD[((String, String), Iterable[Array[String]])] =  
    fineGrainedChanges
(*@\lnote@*)    .groupBy(entry => (entry(EMAIL), entry(PROJECT)))

// aggregate per project
val perProject: RDD[(String, Iterable[Array[String]])] =
    fineGrainedChanges
(*@\lnote@*)    .groupBy(entry => entry(PROJECT))
\end{annotatedcsource}
\vspace{1em}
\end{minipage}

\begin{minipage}{0.95\linewidth}
    \begin{lstlisting}[label={lst:fineGrainedChanges-more-grouping-global},caption={Global operations}]
// unlike previous examples, count() is an "action" which triggers 
// an actual computation that returns a result to the program
// rather than deriving a new RDD
val total: Long = fineGrainedChanges.count()
    \end{lstlisting}
    \vspace{1em}
\end{minipage}

\section{Creating a ground truth dataset}\label{sec:manualLabels}

The first author \textit{manually} classified a randomly sampled set of $\sim$100 commits from each of the studied 11 repositories.
To improve classification quality the projects' issue tracking systems, e.g. JIRA \cite{jira}, was often used. The JIRA contained the tickets occasionally referenced in developers' commits messages (e.g., ``[PRJ-NAME 1234] Fixed some bug''). Such tickets (a.k.a. issues) typically contain additional information about the feature or bug the referencing commit was trying to address. Moreover, tickets sometimes had their own classification labels such as ''feature request``, ''bug``, ''improvement`` etc., but unfortunately they were not very reliable as developers were not always consistent with their labeling (classification). For instance, in some cases bug fixes were labeled as ''improvement``, and while fixing a bug is indeed an improvement, according to the maintenance activities we use \cite{mockus2000identifying}, bug fixes should be classified corrective while improvements should be classified perfective. Some developers used the term ''fix`` even when they referenced feature requests, e.g. ''fixed issue \#N``, where ''issue \#N`` spoke of a new feature or an improvement that did not necessarily report a bug. These observations are consistent with \citet{herzig2013s} who reported that 33.8\% of the bug reports they studied were misclassified.

In cases where the lack of supporting information (e.g., not enough information in the corresponding ticket and / or commit message) prevented us from classifying a certain commit with satisfactory confidence, that commit was discarded from the dataset and replaced by a new one, selected randomly from the same project repository (by re-sampling a commit). 
If we were unable to classify the replacement commit as well, we would repeat this routine until we found a commit that we were able to confidently classify.
Further rules of thumb we used for classifying were as follows:
\begin{itemize}
    \item \textit{Javadoc and comment updates were considered perfective maintenance.} \\ 
    Rational: these changes improve the system.
    \item \textit{Fixing a broken unit test or build was considered corrective maintenance.} \\ 
    Rational: we assume that tests break in the presence of bugs.
    \item \textit{Adding new unit test(s) was considered perfective maintenance.} \\ 
    Rational: we assume that new tests improve coverage. \\
    We conjecture that more often than not, developers who add tests aim to improve system coverage.
    \item \textit{Performance improvements that resulted from an open ticket in the issue tracking system were considered corrective maintenance.} \\ 
    Rational: we assume that tickets that were reported on performance issues resulted from pains on the user side, and addressing these pains is more corrective in nature than perfective.
    \item \textit{Performance improvements that did NOT result from an open ticket in the issue tracking system were considered perfective maintenance.} \\ 
    Rational: we assume that developers may occasionally seize an opportunity to improve code performance, however, if there were no users suffering the problem being fixed, we consider the maintenance to be of a perfective nature, rather than corrective one.
\end{itemize}

We made efforts to avoid class starvation (i.e., not having enough instances of a certain class) by inspecting the proportion of each class within a given sample for a given project. An imbalanced training dataset could substantially degrade models' performance, and in case we detected a considerable imbalance in some project's classes, we added more commits of the starved class from the same project by means of repeatedly sampling and manually classifying commits until a commit of the starved class was found.

To alleviate the challenges involved in reproducing our study we have
made our dataset publicly accessible online \cite{stanislav_levin_2017_835534}. 
This dataset consists of 1151 manually classified commits, 100-115 commits from each of the 11 studied project. Among these commits 43.4\% (500 instances) were corrective, 35\% (404 instances) were perfective, and 21.4\% (247 instances) were adaptive. The commits in this dataset sum up to 33,149 fine-grained source code changes.

\label{sec:manualAgreement}

In order to inspect manual classification agreement, we randomly selected 110 commits out of the 1151 commits, 10 random commits from each of the 11 projects, and had both authors classify it.
At first the agreement stood at 79\%. After discussing the conflicts and sharing the guidelines in more detail, the agreement level rose to 94.5\%. According to the one sample proportion test \cite{altman1990practical}, the error margin for our observed agreement level was 4.2\%, and the estimated asymptotic 95\% confidence interval was [90.3\%, 98.7\%]. 
This indicates that both authors were in agreement about the labels for the vast majority of cases once they employed the same guidelines (see \Cref{sec:manualLabels}).
Regarding some of the commits, no consensus was reached.
Consider a commit with the following message: 
\textit{``add hasSingleArrayBackingStorage allow for- optimization only when there really is a single array, and not when there is- a multi dimensional one''}. One of the annotators had labeled it ``Corrective'', assuming this commit fixed a bug, while the other had labeled it ``Perfective'' assuming this was an optimization which improved performance but did not necessarily fix a known bug. Since there was no JIRA ticket associated with this commit it was difficult to ascertain which label is more plausible. 
Similarly, consider a commit with the message: \textit{``Timeouts for row lock and scan should- be separate''}. Based on the message, this commit could be considered any of the maintenance activities, it could be fixing a bug, improving design (by separating concerns) or adding a new feature (e.g., allowing different timeouts for lock and scan). In this particular case, the referenced JIRA ticket indicated it was an ``improvment'' and thus ``Perfective'', but had it not been for the JIRA ticket it would have been quite challenging to determine the associated maintenance activity.

\section{Commit Classification Models}
\label{modelTypes}
We performed our statistical computations in the R statistical environment \cite{R}, where we extensively used the R caret package \cite{rCaretHome, rCaretDoc} for the purpose of model training and evaluation.

We split the labeled dataset into a training dataset and a test dataset, 85\% and 15\% respectively, in order to have the test dataset completely isolated from any training procedures. The split was performed by using R's \textit{createDataPartition} function \cite{createDataPartition-r}, with the percentage of data that goes to training set to $0.85$. The createDataPartition function uses random sampling within the labels (Corrective, Perfective, Adaptive) in an attempt to balance the class distributions within the splits, see also \Cref{tbl:train-test-dataset-classes} for a detailed description of the train and test splits.

\begin{table}[H]
\center
\renewcommand{\arraystretch}{1.5}
\caption{Number of instances per class in the train and test datasets in our study\protect\footnotemark}
\label{tbl:train-test-dataset-classes}
    \begin{tabular}{|l|c|c|c|}
    \hline
    \rowcolor{lightgray} \textbf{Dataset} & \textbf{Corrective} & \textbf{Perfective} & \textbf{Adaptive} \\ \hline
    Train \small{(979/1151 instances)} & 425 & 344 & 210\\ \hline
    Test \small{(172/1151 instances)} & 75 & 60 & 37 \\ \hline
    \end{tabular}
\end{table}

\begin{table}[H]
\center
\renewcommand{\arraystretch}{1.5}
\caption{Number of commits per studied project in the \textbf{test} dataset}
\label{tbl:test-dataset-projects}
    \begin{tabular}{|l|c|}
    \hline
    \rowcolor{lightgray} \textbf{Project} & \textbf{Commits in test dataset} \\\hline
    RxJava	& 14 \\ \hline
    Restlet &	17 \\ \hline
    Drools	& 16 \\ \hline
    HBase	& 17 \\ \hline
    Spring Framework	& 16 \\ \hline
    OrientDb	& 14 \\ \hline
    Hadoop	& 13 \\ \hline
    Camel	& 13 \\ \hline
    Elasticsearch	& 15 \\ \hline
    Kotlin	& 17 \\ \hline
    Intellij Community Edition & 20 \\ \hline
    \end{tabular}
\end{table}

\footnotetext{The entire labeled dataset, consisting of 1151 labeled commits, is publicly available at \url{https://doi.org/10.5281/zenodo.835534}, see also \citet{stanislav_levin_2017_835534}.}

The model training phase consists of using 5 time repeated 10-fold validation for each compound model on the training dataset (which boils down to performing a 10-fold cross validation process 5 different times and averaging the results). Then, the trained models were evaluated using the test dataset - the 15\% split that did not take part in the model training process.

\subsection{Utilizing word frequency analysis}

First we classified the test dataset (the 15\% of the entire labeled dataset) using a naive method to set an initial baseline. 
The naive method is based on a classification technique described in our previous work \cite{levinIcsme2016}, and consists of searching for pre-defined words (see \Cref{tab:classificationWords}), and assigning the most frequent class (i.e., corrective) in case none of the keywords were present in the commit message, see \Cref{tab:veryNaiveClassification} for more details.
Assigning the most frequent class to an instance is far from ideal, however, when models find no features to rely on, using the overall distribution of the training dataset is a common technique (also called 'No Information Rate', see \Cref{sec:statMethods}).

\begin{table}[H]
\renewcommand{\arraystretch}{2}
\caption{Stemmed keywords used by the ``naive method'' as described in \citep{levinIcsme2016}}
\label{tab:classificationWords} 
    \begin{tabular}{|c|p{10cm}|}
    \hline
        \cellcolor{lightgray} \textbf{Corrective} & \textit{fix,
          esolv,
          clos,
          handl,
          issue,
          defect,
          bug,
          problem,
          ticket} \\
    \hline
        \cellcolor{lightgray} \textbf{Perfective} & \textit{refactor,
          re-factor,
          reimplement,
          re-implement,
          design,
          replac,
          modify,
          updat,
          upgrad,
          cleanup,
          clean-up} \\
    \hline
        \cellcolor{lightgray} \textbf{Adaptive} &  \textit{add,
          new,
          introduc,
          implement,
          extend,
          feature,
          support} \\
    \hline
    \end{tabular}
\end{table}

\begin{table}[H]
    \center
    \renewcommand{\arraystretch}{1.5}
        \caption{Naive model's confusion matrix}        
        \label{tab:veryNaiveClassification}
        \begin{tabular}{|c|c|c|c|}        
        \hline
        \rowcolor{lightgray} \backslashbox{\Large classified as}{\Large true class} & \textbf{Adaptive} &  \textbf{Corrective} &  \textbf{Perfective} \\
        \hline
        \textbf{Adaptive} &  \textbf{18}  & 2 & 16 \\
        \hline
        \textbf{Corrective}  &   18 & \textbf{72} & 37 \\
        \hline              
        \textbf{Perfective} &     1 &  1 & \textbf{7} \\
        \hline        
        \Xhline{1pt}                    
        \multicolumn{1}{r|}{\textbf{Recall:}} & 48\% & 96\% & 11\% \\ \cline{2-4}
        \multicolumn{1}{r|}{\textbf{Precision:}} & 50\% & 56\% & 77\% \\ \cline{2-4}
        \multicolumn{1}{r|}{\textbf{Accuracy:}} & \multicolumn{3}{c|}{56\%} \\ \cline{2-4}
        \multicolumn{1}{r|}{\textbf{Kappa:}} & \multicolumn{3}{c|}{29\%} \\ \cline{2-4}
        \multicolumn{1}{r|}{\textbf{F1 Score (micro-averaged):}} & \multicolumn{3}{c|}{0.56} \\ \cline{2-4}
        \multicolumn{1}{r|}{\textbf{F1 Score (macro-averaged):}} & \multicolumn{3}{c|}{0.46} \\
        \cline{2-4}
        \multicolumn{1}{r|}{\textbf{No Information Rate (NIR):}} & \multicolumn{3}{c|}{ 43\% } \\ \cline{2-4}
        \multicolumn{1}{r|}{\textbf{P-Value [Accuracy $>$ NIR]:}} & \multicolumn{3}{c|}{ 0.0005 }  \\ \cline{2-4}
        \end{tabular}
\end{table}    

The results showed that 34.8\% of the commits in the test dataset (60 commits) did not have any of the keywords present in their commit message, and were therefore automatically classified corrective. In addition, the low recall of the perfective class was particularly notable, as opposed to the high recall of the corrective class (which accounts for most of the commits in the classified dataset).
The noticeable difference between the micro-averaged and macro-averaged F1 scores, 0.56 vs. 0.46 respectively, also indicates that the current model (based on the naive method) does not perform equally well for all classes.

The high percentage of commits without any keywords prompted us to try to fine-tune the keywords we were searching for.
We performed an additional experiment using the same classification method, only this time the keywords were obtained by employing a word frequency analysis and normalization for the commit messages. This time 28\% of the commits did not have any of the keywords present in their commit message. 
These findings led us to believe that the high number of commit messages containing none of the keywords could be playing a significant role in determining the overall classification quality.

\subsection{Utilizing source code changes}\label{sec:models}

Techniques for dealing with missing values in classification problems are broadly covered by \citet{saar2007handling}, who describe two common methods used to overcome such issues: 
\begin{enumerate*}
    \item imputation, where the missing values are estimated from the data that are present, and
    \item reduced-feature models, which employ only those features that will be known for a particular test case (i.e., only a subset of the features that are available for the entire training dataset), so that imputation is not necessary.
\end{enumerate*}
Since our dataset consists of two different data types, keywords and source code changes, we use reduced-feature models, which are reported to outperform imputation and represent our use-case more naturally.
In addition, since the missing feature patterns in our dataset are known in advance, i.e., given a commit only the keywords can be missing, its source code changes are always present, we can pre-compute and store two models; one to be used when all features are present (keywords $+$ source code changes), and the other when only a subset is available (source code changes only).
We define the notion of a \textbf{compound} model (similarly to the ``classifier lattice'' described by \citeauthor{saar2007handling}) which uses two separate models for classifying commits with, and without (pre-defined) keywords in their commit message.
The classify routine of the compound model is pseudo-coded in \Cref{compound-classify-cod}.

\begin{minipage}{0.95\linewidth}
\begin{annotatedcsource}{compound-classify-cod}{The compound model's classify routine}
classify(commit) {
    if(hasKeywords(commit.comment)) { 
(*@\lnote@*)        return classifyWith($\mathit{model}_{KW}$,commit)
    } else {
(*@\lnote@*)        return classifyWith($\mathit{model}_{\overline{KW}}$,commit)
    }
}
\end{annotatedcsource}
\vspace{1em}
\end{minipage}

\noindent Given a commit $C$, the compound model first checks if $C$'s commit message has any keywords, if so, the model defined as $\mathit{model}_{KW}$ is used to classify $C$ (see~bookmark~\lnnum{1}~in~\Cref{compound-classify-cod}), otherwise (i.e., no keywords found in $C$'s commit message), the model defined as $\mathit{model}_{\overline{KW}}$ is used to classify $C$ (see~bookmark~\lnnum{2}~in~\Cref{compound-classify-cod}).
Each of the models $\mathit{model}_{KW}$ and $\mathit{model}_{\overline{KW}}$ may or may not be a reduced-feature model, depending on whether it employs the full set of features (both keywords and source code changes), or only a subset of it (either keywords or source code changes).\\
We define $\mathit{model}_{\overline{KW}}$ and $\mathit{model}_{KW}$ to be one of the following model types:

\begin{itemize}
    \item \underline{Keywords} model, which relies solely on keywords to classify commits. 
        \label{keywordsExtraction}
        The features used by this model are keywords obtained by performing the following transformations on the commit message field:
        
        \begin{enumerate}
            \item Stripped special characters 
            \item Made lower case (case-folding)
            \item Stripped English stopwords
            \item Stripped punctuation
            \item Striped white-spaces
            \item Performed stemming
            \item Adjusted frequencies so that each comment can contribute a given word only once
            \item Stripped custom words such as developer names, projects names, VCSs lingo (e.g., head, patch, svn\footnote{Subversion is commonly abbreviated to SVN after its command name \texttt{svn}.}, trunk, commit), domain specific terms (e.g., http, node, client):
            {\small ''patch``, ''hbase``, ''checksum``, ''code``, ''version``, ''byte``, ''data``, ''hfile``, ''region``, ''schedul``, ''singl``, ''can``, ''yarn``, ''contribut``, ''commit``, ''merg``, ''make``, ''trunk``, ''hadoop``, ''svn``, ''ignoreancestri``, ''node``, ''also``, ''client``, ''hdfs``, ''mapreduc``, ''lipcon``, ''idea``, ''common``, ''file``, ''ideadev``, ''plugin``, ''project``, ''modul``, ''find``, ''border``, ''addit``, ''changeutilencod``, ''clickabl``, ''color``, ''column``, ''cach``, ''jbrule``, ''drool``, ''coprocessor``, ''regionserv``, ''scan``, ''resourcemanag``, ''cherri``, ''gong``, ''ryza``, ''sandi``, ''xuan``, ''token``, ''contain``, ''shen``, ''todd``, ''zhiji``, ''tan``, ''wangda``, ''timelin``, ''app``, ''kasha``, ''kashacherri``, ''messag``, ''spr``, ''camel``, ''http``, ''now``, ''class``, ''default``, ''pick``, ''via``.}

            \item We then selected the 10 most frequent words from each of the three maintenance activities in the test dataset:                             
                \begin{itemize}
                    \item Corrective:
                        \begin{enumerate*}[label={(\arabic*)},before=\itshape,font=\normalfont]
                            \item fix
                            \item test
                            \item issu
                            \item use
                            \item fail
                            \item bug
                            \item report
                            \item set 
                            \item error 
                            \item npe
                        \end{enumerate*}
                    \item Perfective:
                        \begin{enumerate*}[label={(\arabic*)},before=\itshape,font=\normalfont]
                            \item test
                            \item remov
                            \item use
                            \item fix
                            \item refactor
                            \item method
                            \item chang 
                            \item add
                            \item improv
                            \item new                           
                        \end{enumerate*}                    
                    \item Adaptive:
                        \begin{enumerate*}[label={(\arabic*)},before=\itshape,font=\normalfont]
                            \item support
                            \item add
                            \item implement
                            \item new
                            \item allow
                            \item use
                            \item method
                            \item test
                            \item set
                            \item chang
                        \end{enumerate*}
                \end{itemize}                        
                        
            It can be seen that some of the words (as obtained by our commit message word frequency analysis) overlap between maintenance activities. The words \textit{''test``} and \textit{''use``} appear in all three maintenance activities; the word \textit{''fix``} appears in both the corrective and perfective maintenance activity; the words \textit{''method``}, \textit{''chang``}, \textit{''add``} and \textit{''new``} appear both in the perfective and adaptive maintenance activities; and the word \textit{''set``} appears both in the corrective and adaptive maintenance activities.
            These word overlaps may indicate that keywords alone are insufficient to accurately classify commits into maintenance activities, and need to be augmented with additional information in order to improve classification accuracy.
            
            For the purpose of building the Keywords model type, we remove multiple occurrences of the same word (so that each word appears only once in the combined list) and remain with the following set of words:
            \begin{enumerate*}[label={(\arabic*)},font=\normalfont]
                \item add
                \item allow
                \item bug
                \item chang
                \item error
                \item fail
                \item fix
                \item implement
                \item improv
                \item issu
                \item method
                \item new
                \item npe
                \item refactor
                \item remov
                \item report
                \item set
                \item support
                \item test
                \item use.
            \end{enumerate*}        
        \end{enumerate}
    \item \underline{(Source Code) Changes} based model, which relies solely on source code changes to classify commits.
        The features used by this model are source code change types \cite{fluri2006classifying} obtained by distilling commits, as described earlier in this section.
    \item \underline{Combined} (Keyword + Source Code Change Types) model, which uses both keywords and source code change types to classify commits.
        The features used by this type of models consist of both keywords and source code change types. 
\end{itemize}

A word-cloud visualization of the keyword distribution in each of the maintenance activities can be found in \Cref{fig:wordcloud-corrective}, \Cref{fig:wordcloud-perfective}, \Cref{fig:wordcloud-adaptive}. A summary of the model components can be found in \Cref{internalComponentTypes}.

\begin{table}[ht]    
    \center
    \renewcommand{\arraystretch}{2}
    \caption{Reduced-feature model components}
    \label{internalComponentTypes}
    \begin{tabular}{|c|c|}        
    \hline
    \rowcolor{lightgray} \textbf{Model Type} &  \textbf{Model Features} \\
    \hline
    Keywords &  Words \\
    \hline
    Changes  & Fine-grained Source Code Change Types  \\
    \hline              
    Combined & Words + Fine-grained Source Code Change Types  \\
    \hline
    \end{tabular}    
\end{table}

For example, a commit where two methods were added (fine-grained source code change type ''additional\_functionality``), and one statement was updated (fine-grained source code change type ''statement\_updated``) and has a commit message that says ''Refactored blob logic into separate methods`` will be treated differently by each of the model types indicated in \Cref{internalComponentTypes}.\\
The Keywords model extracts features represented by tuples of size 20, and given the commit above would extract the following features:
$\overbrace{(0\dots1\dots1\dots0)}^\text{20}$ with ``1'' in the coordinates that represent the words \textit{''refactor``} and \textit{''method``}. The count of each keyword is at most one, i.e., duplicate keywords are counted only once. Source code changes are ignored, since the Keywords model type does not consider source code changes.\\
The Changes model extracts features represented by tuples of size 48 (since there are 48 different source code change types), and given the commit above would extract the following features:
$\overbrace{(0\dots2\dots1\dots0)}^\text{48}$ with ''2`` in the coordinate that represents the fine-grained source code change type \textit{''additional\_functionality``} and ``1'' in the coordinate that represents \textit{''statement\_updated``}. In contrast to the case of the Keywords model, all occurrences of every fine-grained source code change type are counted in. Keywords in the commit message are ignored, since the Changes model type does not consider keywords.\\
The Combined model extracts features represented by tuples of size 68 ($=$ 48 fine-grained source code change types + 20 keywords), and given the commit above would extract the following features:
$\overbrace{(\underbrace{0\dots1\dots1\dots0}_\text{20}\underbrace{\dots0\dots2\dots1\dots0}_\text{48})}^\text{68}$, with ''2`` in the coordinate that represents the fine-grained source code change type \textit{''additional\_functionality``}, and ''1`` in the coordinates that represent the fine-grained source code change type \textit{''statement\_updated``}, the keyword \textit{''refactor``}, and the keyword \textit{''method``}. The Combined model type captures both keywords and fine-grained source code change types - hence its name.

In the next sections we evaluate and compare different compound models by considering the different combinations of their $\mathit{model}_{KW}$ and $\mathit{model}_{\overline{KW}}$ model components.
The evaluation process consists of the following steps:
\begin{enumerate}
    \item Select the model component $\mathit{model}_{KW}$
    \item Select the model component $\mathit{model}_{\overline{KW}}$    
    \item Select an underlying classification algorithm for the compound model, which determines the algorithm to be used by each of the model components $\mathit{Model}_{KW}$ and $\mathit{Model}_{\overline{KW}}$ (J48, GBM, or RF, see also \Cref{sec:methods}).        
\end{enumerate}

\section{Evaluation}\label{sec:evaluaton}

\setlength\emergencystretch{\hsize}

We describe an exhaustive set of combinations for selecting the pair of $(\mathit{Model}_{KW}, \mathit{Model}_{\overline{KW}})$ models in \Cref{tab:model_combos}, where the pairs can be one of the three model types defined in \Cref{internalComponentTypes}.
Each row in \Cref{tab:model_combos} represents a compound model, defined by the selection of $(\mathit{Model}_{KW}, \mathit{Model}_{\overline{KW}})$. The classification accuracy and Kappa achieved by a given compound model are reported in the corresponding Accuracy and Kappa columns.
The best performing compound model for each classification algorithm is highlighted in lime-green, and the keywords based model (where both $\mathit{Model}_{KW}$ and , $\mathit{Model}_{\overline{KW}}$ are of the Keywords model type) is highlighted in orange so that it can be easily compared to compound models that utilize fine-grained source code changes.

\begin{table}[H]
    \renewcommand{\arraystretch}{1.3}
    \centering
    \caption{Training dataset compound models performance}
    \label{tab:model_combos}
    \begin{tabular}{|p{0.13\linewidth}|c|c|c|c|}
        \hline
        \rowcolor{lightgray} \centering \textbf{Alg.}  & $\mathit{Model}_{KW}$ & $\mathit{Model}_{\overline{KW}}$ & \textbf{Accuracy} & \textbf{Kappa} \\
        \hline
         \centering  \multirow{9}{*}{J48}  & \multicolumn{2}{c|}{Combined}                                                      &      69.0\% & 51.7\% \\
         \cline{2-5}
         & Combined     &       Keywords                                                                            &        67.7\% & 50.2\%  \\
         \cline{2-5}
         & Combined     &       Changes                                                                             &         69.2\% & 51.9\% \\
         \cline{2-5}
         \Xcline{2-5}{1.5pt}
         &  Keywords    &     Combined                                                        &    69.8\% &  53\% \\
         \cline{2-5}
         & \multicolumn{2}{c|}{\cellcolor{orange} Keywords}                                                                        &       \cellcolor{orange} 68.5\% & \cellcolor{orange} 51.5\% \\
         \cline{2-5}
         & \cellcolor{lime} Keywords    &  \cellcolor{lime} Changes  &    \cellcolor{lime}  69.9\% & \cellcolor{lime} 53.2\% \\
         \cline{2-5}
         \Xcline{2-5}{1.5pt}
         & Changes  &       Combined                                                                                &         48.7\% & 20.1\% \\
         \cline{2-5}         
         & Changes  &       Keywords                                                                            &       47.4\% & 17.2\%  \\
         \cline{2-5}
         & \multicolumn{2}{c|}{Changes}                                                                         &       48.8\% & 18.6\% \\
        \Xhline{3pt}

         \multirow{9}{*}{\parbox{\linewidth}{\centering GBM}}  & \multicolumn{2}{c|}{\cellcolor{lime}Combined}  & \cellcolor{lime} 72.0\% &  \cellcolor{lime} 56.2\% \\
         \cline{2-5}
         & Combined   &         Keywords                                                                            &      69.0\% & 51.8\% \\
          \cline{2-5}
         &   Combined   &     Changes                                                                               &   72.0\% & 55.9\% \\
          \cline{2-5}
          \Xcline{2-5}{1.5pt}
         & Keywords   &        Combined                                                                             &       71.6\% & 56.0\% \\
          \cline{2-5} 
          
         & \multicolumn{2}{c|}{\cellcolor{orange} Keywords}                                                                        & \cellcolor{orange} 68.5\% & \cellcolor{orange} 51.4\% \\         
          \cline{2-5}
         & Keywords    &       Changes                                                                          &       71.5\% & 55.6 \\
          \cline{2-5}
          \Xcline{2-5}{1.5pt}
         & Changes  &       Combined                                                                                &       54.1\% & 26.9\% \\
          \cline{2-5}
         & Changes   &      Keywords                                                                            &      51.0\% & 22.4\% \\
          \cline{2-5}
         & \multicolumn{2}{c|}{Changes}                                                                         &       54.3\% & 26.9\% \\
         \Xhline{3pt}

        \multirow{9}{*}{\parbox{\linewidth}{\centering RF}}                     &  \multicolumn{2}{c|}{Combined}    &       73.1\% & 57.8\% \\
          \cline{2-5}
          & Combined     &       Keywords                                                                           &      69.5\% & 52.6\% \\
          \cline{2-5}
          & Combined     &       Changes                                                                            &      71.9\% & 55.7\% \\
          \cline{2-5}
          \Xcline{2-5}{1.5pt}
          & Keywords    &       Changes                                                                         &       72.2\% & 56.4\% \\
          \cline{2-5}
          & \cellcolor{lime} Keywords    &   \cellcolor{lime}    Combined       &    \cellcolor{lime}  73.6\% & \cellcolor{lime} 58.9\% \\
          \cline{2-5}
          & \multicolumn{2}{c|}{\cellcolor{orange} Keywords}                                                                       & \cellcolor{orange} 69.8\% & \cellcolor{orange} 53.4\% \\
          \cline{2-5}        
          \Xcline{2-5}{1.5pt}
          & Changes  &       Combined                                                                               &       54.5\% & 26.6\% \\
          \cline{2-5}
          & Changes  &       Keywords                                                                           &       50.6\% & 21.1\% \\
          \cline{2-5}
          & \multicolumn{2}{c|}{Changes}                                                                        &       52.9\% & 23.4\% \\
    \hline         
    \end{tabular}
\end{table}

Following our main research questions (see \Cref{sec:intro}), the accuracy and Kappa results for each compound model during the training (see \Cref{tab:model_combos}) reveal that the compound models that use either \fbox{$ \displaystyle \mathit{Model}_{\overline{KW}} = \mathit{Combined}$} or \fbox{$ \displaystyle \mathit{Model}_{\overline{KW}} = \mathit{Changes}$} achieve higher accuracy and Kappa when compared to models with the same $\displaystyle \mathit{Model}_{KW}$ component but  with ${\mathit{Model}_{\overline{KW}} = \mathit{Keywords}}$, regardless of the underlying classification algorithm (J48, GBM or RF). This comes as no surprise, as one could expect keyword based models would have trouble accurately classifying commits that do not have any keywords in their commit message. \Cref{tab:model_combos} also reveals that models that rely solely on commit messages have higher accuracy and Kappa than models that rely solely on fine-grained source code changes (under all three algorithms).

Further accuracy and Kappa statistics pertaining to the training stage of the best performing model for each algorithm can be found in \Cref{resamplesStatsAccuracy} and \Cref{resamplesStatsKappa} respectively. From \Cref{resamplesStatsAccuracy} and \Cref{resamplesStatsKappa} we can learn that during the training stage, the RF model consistently outperforms the J48 and even the GBM model, in both accuracy and Kappa, across all of the cuts: minimum, 1-st quartile (25-th percentile), median, mean, 3-rd quartile (75-th percentile) and maximum. In particular, the minimum accuracy and Kappa of the RF are notably higher than its competitors.

\begin{table}
    \center
    \newcommand\Tstrut{\rule{0pt}{2.6ex}}
    \renewcommand{\arraystretch}{1.5}
    \setlength\extrarowheight{2pt}
    \caption{Training dataset accuracy, best model per algorithm}
    \label{resamplesStatsAccuracy}
    \begin{tabular}{|c|c|c|c|c|c|c|}
    \hline
    \rowcolor{lightgray} Alg. & Min. & 1-st Q. & Median & Mean & 3-rd Q. &  Max. \\
    \hline
    J48 & 60.8\% & 66.4\% & 70.1\% & 69.9\% & 73.4\% & 80.6\% \\
    \hline
    GBM & 60.8\% & 69.2\% & 72.1\% & 72.0\% & 75.2\% & 80.8\% \\
    \hline              
    \rowcolor{lime} RF &  65.6\% &  70.4\%  & 73.4\% & 73.6\%  & 76.6\% & 82.8\% \\
    \hline
    \end{tabular}    
\end{table}

\begin{table}
    \center
    \renewcommand{\arraystretch}{1.5}
    \setlength\extrarowheight{2pt}
    \caption{Training dataset Kappa, best model per algorithm}
    \label{resamplesStatsKappa}
    \begin{tabular}{|c|c|c|c|c|c|c|}
    \hline
    \rowcolor{lightgray} Alg. & Min. & 1-st Q. & Median & Mean & 3-rd Q. &  Max. \\
    \hline
    J48 & 38.4\% & 47.9\% & 53.4\% & 53.2\% & 58.8\% & 69.7\% \\
    \hline
    GBM & 38.3\% & 51.8\% & 56.9\% & 56.2\% & 60.6\% & 70.0\% \\
    \hline              
    \rowcolor{lime} RF &  45.5\% & 54.1\% & 58.6\% & 58.9\% & 63.3\% & 73.5\% \\
    \hline
    \end{tabular}    
\end{table}

\begin{figure}[H]    
    \centering
    \captionsetup{belowskip=12pt}
    \caption{Training dataset accuracy and Kappa, best model by algorithm}
    \label{fig:comparingModels}    
    \includegraphics[width=0.7\linewidth]{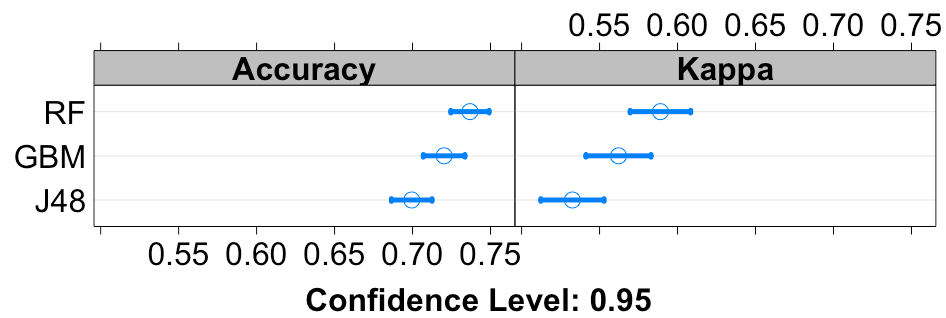}
\end{figure}

A comparison between the best compound models from each of the underlying classification algorithm category can be found in \Cref{fig:comparingModels}.
The top performing models were then used to classify the test dataset, consisting of 15\% of the entire labeled dataset, see \Cref{tab:testsetResults}.
The ultimate winner was the RandomForest compound model with $\mathit{Model}_{KW}=\mathit{Keywords}$ and $\mathit{Model}_{\overline{KW}}=\mathit{Combined}$.
A detailed confusion matrix for this champion  model can be found in \Cref{tab:bestClassification}.

\begin{table}[H]
    \renewcommand{\arraystretch}{1.5}
    \centering
    \caption{Test dataset classification performance}
    \label{tab:testsetResults}
    \begin{tabular}{|c|c|c|c|c|}
        \hline
        \rowcolor{lightgray} \textbf{Algorithm}  & $\mathit{Model}_{KW}$ & $\mathit{Model}_{\overline{KW}}$ & \textbf{Accuracy} & \textbf{Kappa}  \\
        \hline
         J48 & Keywords & Changes & 70 \% & 53\% \\
         \hline
         GBM & \multicolumn{2}{c|}{Combined}  & 72 \% & 57 \% \\
         \hline
         \rowcolor{lime} RF & Keywords & Combined & 76 \% & 63 \% \\
         \hline
     \end{tabular}
 \end{table}

\begin{table}[H]
    \center
    \renewcommand{\arraystretch}{1.5}
        \caption[RF based Keywords-Combined compound model's confusion matrix, test dataset]{RF based Keywords-Combined compound model's confusion matrix for the test dataset}
        \label{tab:bestClassification}
        \begin{tabular}{|c|c|c|c|}        
        \hline
        \rowcolor{lightgray} \backslashbox{\Large classified as}{\Large true class} & \textbf{Adaptive} &  \textbf{Corrective} &  \textbf{Perfective} \\
        \hline
        \textbf{Adaptive} &  \textbf{28}  & 5 & 5 \\
        \hline
        \textbf{Corrective}  &   6 & \textbf{63} &  14 \\
        \hline              
        \textbf{Perfective} &     3 &  7  & \textbf{41} \\
        \hline
        \Xhline{1pt}
        \multicolumn{1}{r|}{\textbf{Recall:}} & 75\% & 84\% & 68\% \\ \cline{2-4}        
        \multicolumn{1}{r|}{\textbf{Precision:}} & 73\% & 75\% & 80\% \\ \cline{2-4}        
        \multicolumn{1}{r|}{\textbf{Accuracy:}} & \multicolumn{3}{c|}{76\%} \\\cline{2-4}
        \multicolumn{1}{r|}{\textbf{Kappa:}} & \multicolumn{3}{c|}{63\%} \\\cline{2-4}
        \multicolumn{1}{r|}{\textbf{F1 Score (micro-averaged):}} & \multicolumn{3}{c|}{0.76} \\\cline{2-4}
        \multicolumn{1}{r|}{\textbf{F1 Score (macro-averaged):}} & \multicolumn{3}{c|}{0.76}\\\cline{2-4}
        \multicolumn{1}{r|}{\textbf{No Information Rate (NIR):}} & \multicolumn{3}{c|}{43\%} \\\cline{2-4}
        \multicolumn{1}{r|}{\textbf{P-Value [Accuracy $>$ NIR]:}} & \multicolumn{3}{c|}{$<2e^{-16}$ } \\ \cline{2-4}         
        \end{tabular}    
\end{table}    

\begin{figure}[H]
    \centering
    \caption{A J48 Keywords model type (''a`` stands for adaptive, ''c`` for corrective, and ''p`` for perfective)}
    \label{fig:j48-tree}
    \includegraphics[height=0.8\textwidth,angle=-90]{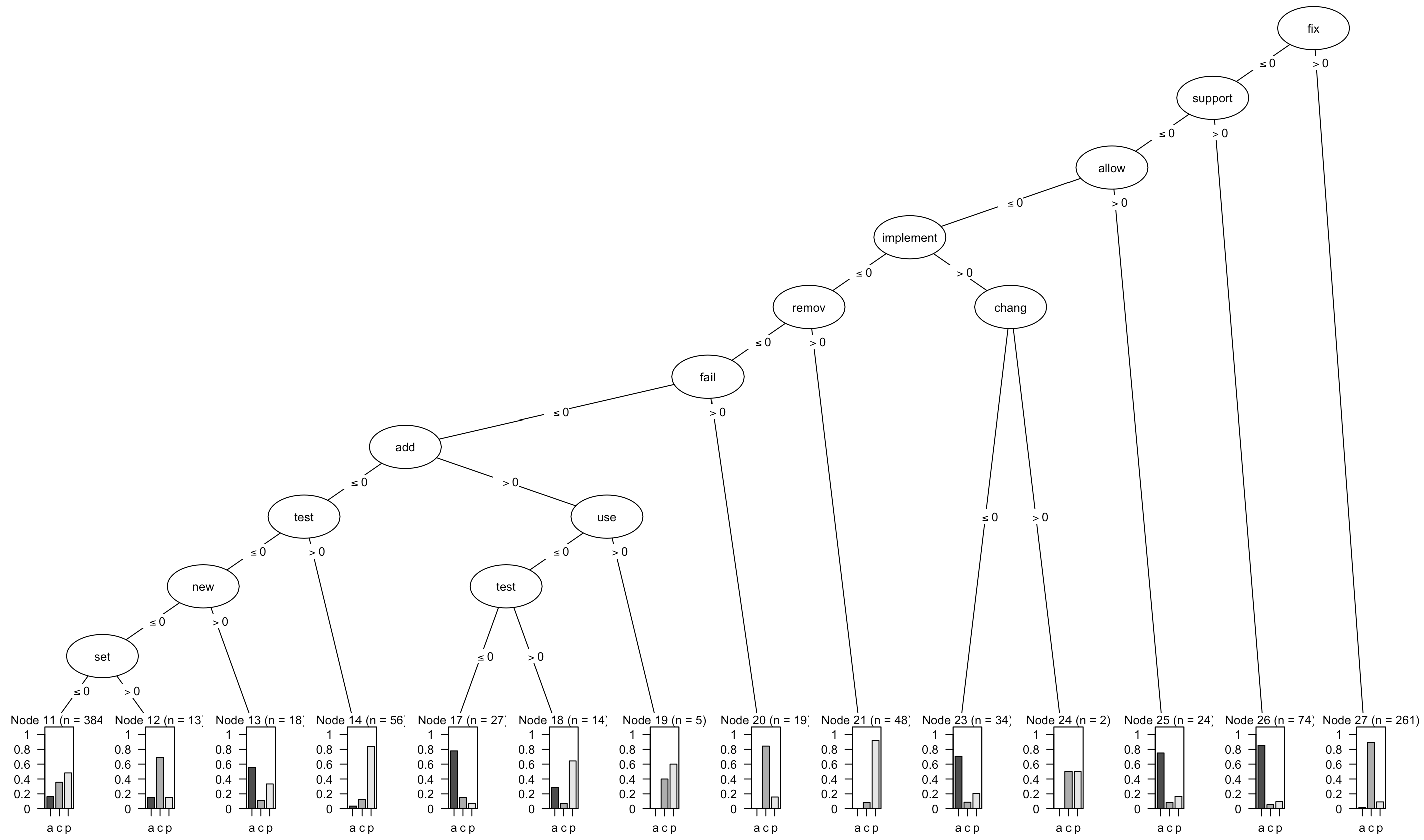}
\end{figure}

The decision tree built by the J48 algorithm for our keyword based model (see \Cref{fig:j48-tree}) provides some interesting insights regarding its classification process.
The word ''fix`` is the single most indicative word of corrective commits, which aligns well with our intuition, according to which commits that fix faults are likely to include the ''fix`` noun or verb in the commit message. Given that ''fix`` did not appear, the words ''support`` and ''allow`` are most indicative of adaptive commits, presumably these words are used by developers to indicate the support of a new feature, or the fact that something new is now ''allowed`` in the system. 
The combination ''implement chang`` (stemmed), given that ''fix``, ''support`` and ''allow`` did not appear, is very indicative of either perfective or corrective commits, if however, ''implement`` is not accompanied by the word ''chang`` (stemmed), the commit is likely to be adaptive.
The (stemmed) word ''remov``, given that the words ''fix``, ''support``, ''allow`` and ''implement`` did not appear, is very indicative of perfective commits, perhaps because developers often use it to describe a modification where they remove an obsolete mechanism in favor of a new one.

\begin{figure}[H]
    \center
    \caption{Word-cloud for the ``Corrective'' maintenance activity}
    \label{fig:wordcloud-corrective}    
    \vspace{5mm}
    \includegraphics[width=\textwidth]{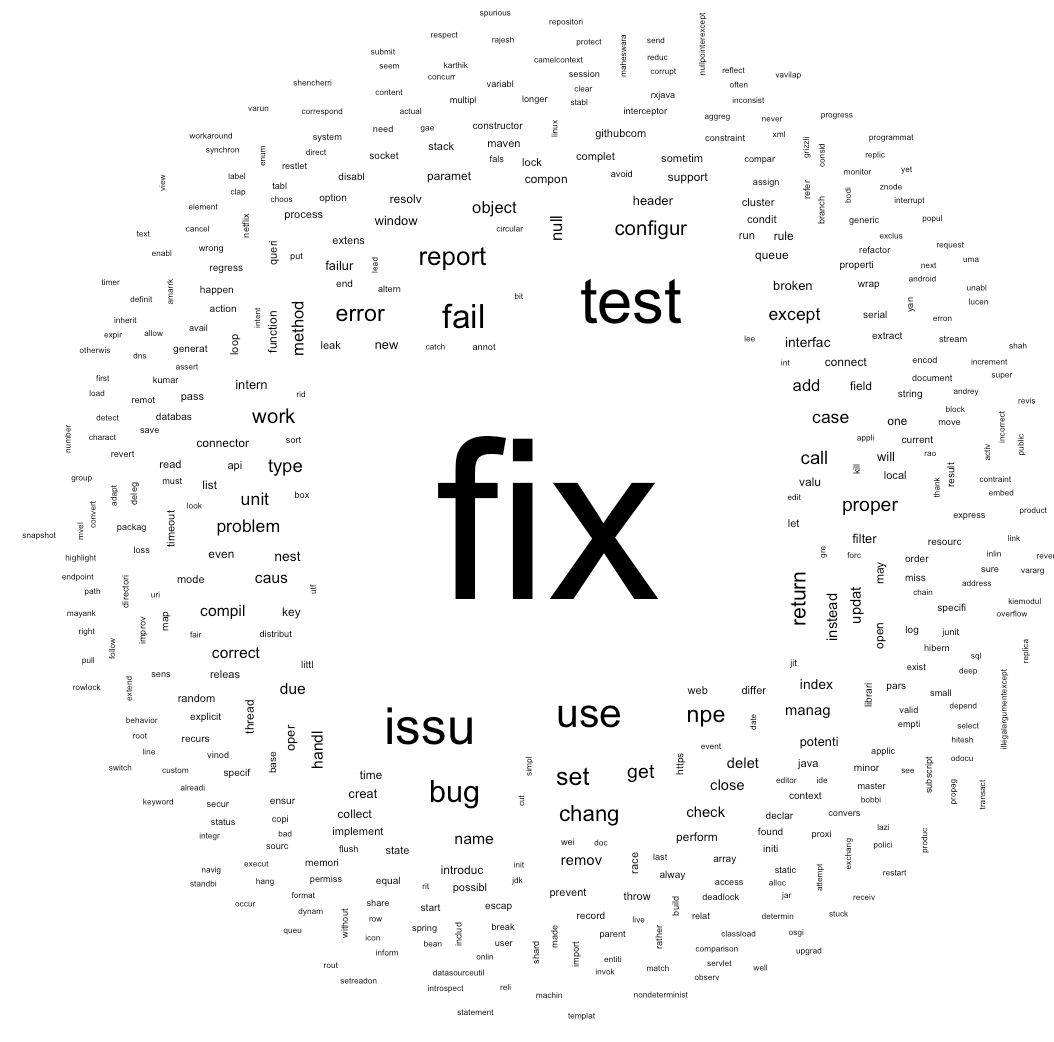}
\end{figure}

\begin{figure}[H]
    \center
    \caption{Word-cloud for the ``Perfective'' maintenance activity}
    \label{fig:wordcloud-perfective}    
    \vspace{5mm}
    \includegraphics[width=\textwidth]{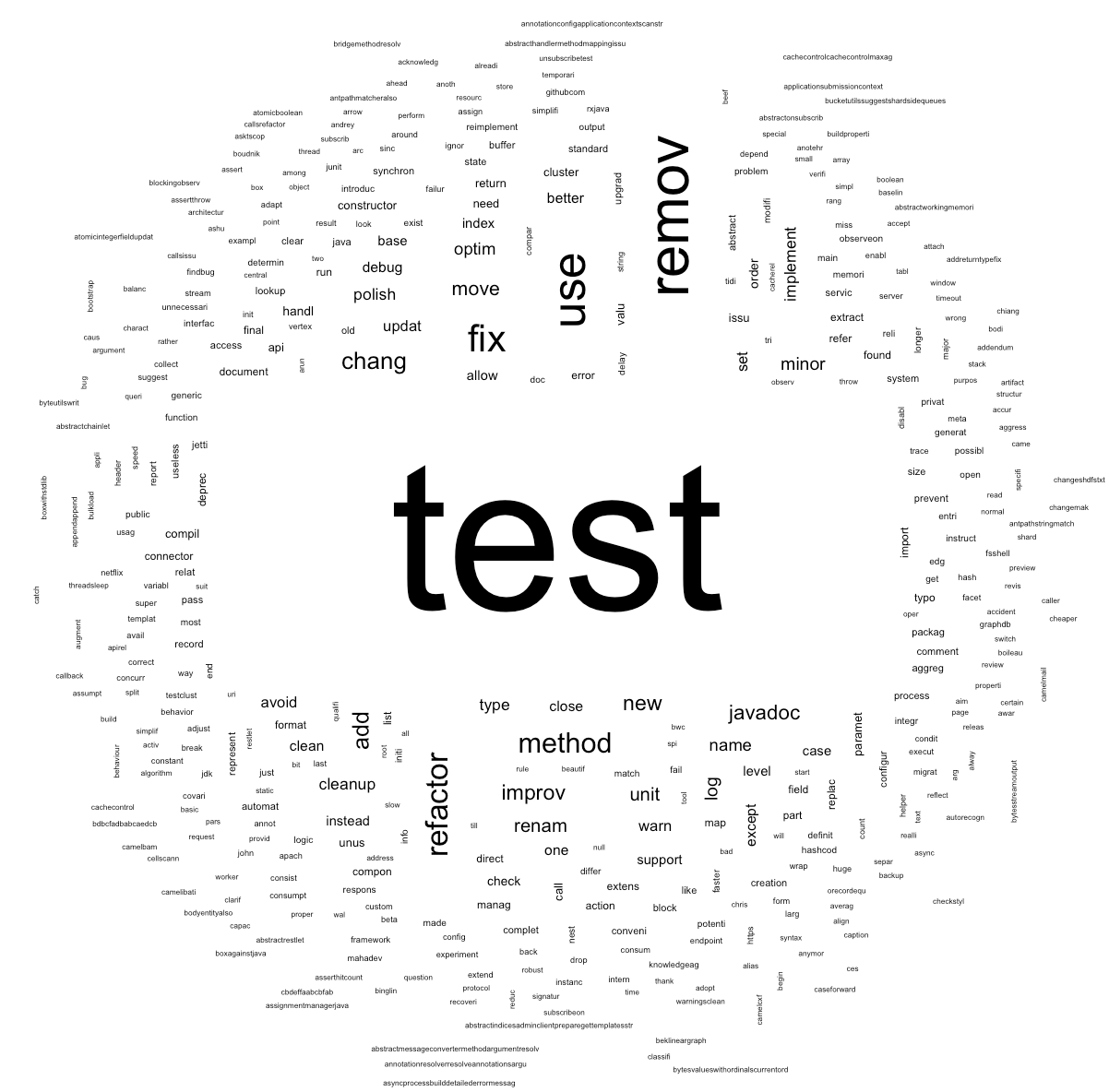}
\end{figure}

\begin{figure}[H]
    \center
    \caption{Word-cloud for the ``Adaptive'' maintenance activity}
    \label{fig:wordcloud-adaptive}    
    \vspace{5mm}
    \includegraphics[width=\textwidth]{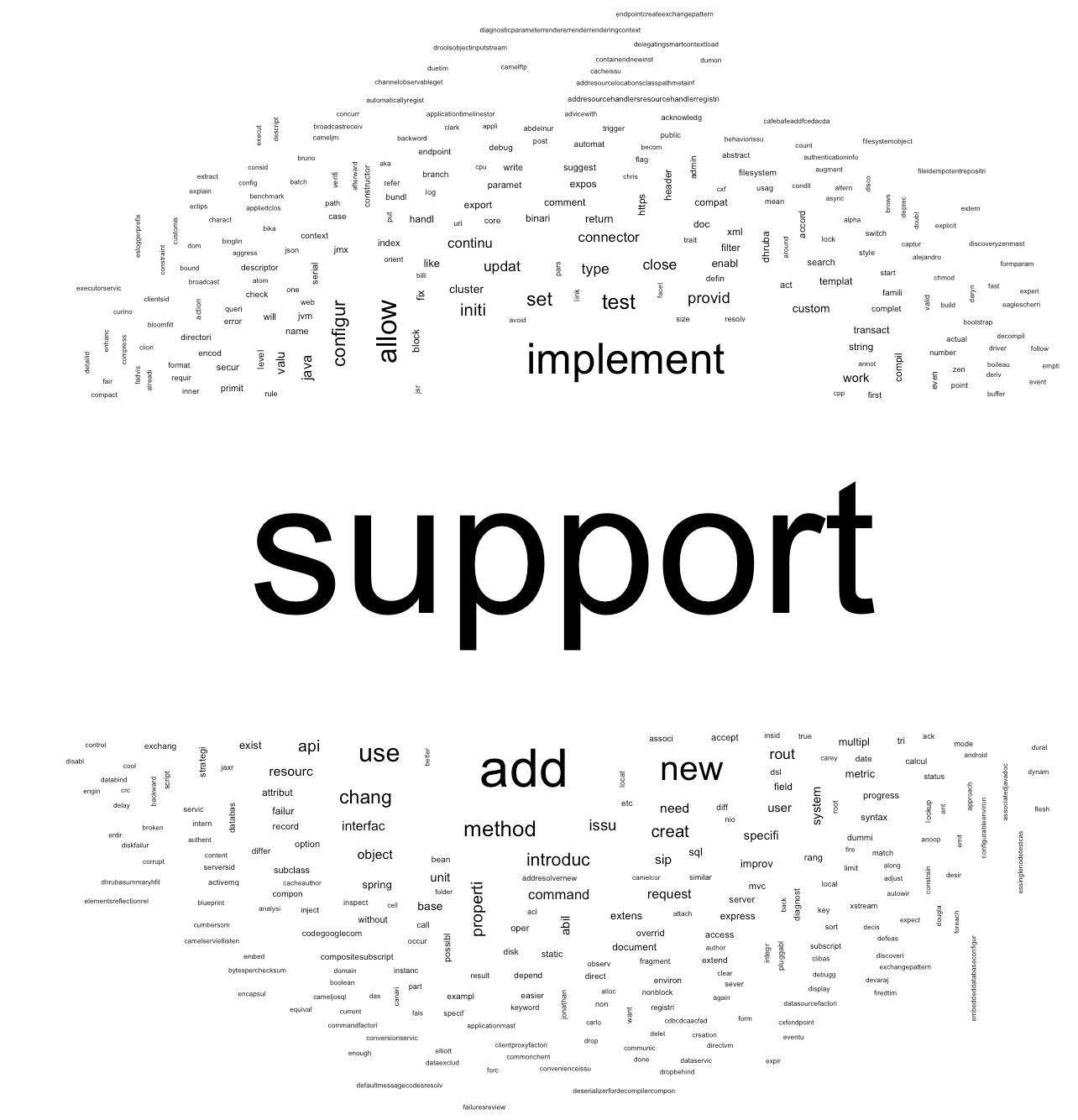}
\end{figure}

We also visualized the keyword frequency in maintenance activities using a word-cloud (see \Cref{fig:wordcloud-corrective}, \Cref{fig:wordcloud-perfective}, \Cref{fig:wordcloud-adaptive}), which revealed that the word ''test`` is particularly common in perfective commits, but is generally common in all three maintenance activity types. The word ''use`` is also common in all three maintenance activity types, but is particularly frequent in the perfective maintenance activity.
The words ''fix``, ''remov`` and ''support`` are quite distinctive of their corresponding maintenance activity types: corrective, perfective and adaptive (respectively). 
The word ''add`` is common in adaptive commits, as well as ''allow``.

Similarly, we visualized the fine-grained source code changes frequencies using a source-code-change-type-cloud which revealed that statement related changes, e.g., ''statement\_insert``, ''statement\_update`` and ''statement\_delete`` are the most common change types in all three maintenance activities (corrective, perfective, adaptive). The fine-grained source code change type ''additional\_functionality`` is common in both perfective and adaptive commits, but less so in corrective commits.


The term-cloud and J48 keyword based decision tree visualizations provide an intuition for why J48 is likely to outperform a simple word-frequency based classification. In contrast to the word-cloud, which provides ''flat`` frequencies, the J48 is capable of capturing information pertaining to the presence of multiple keywords in the same commit message, as indicated by the decision tree.

We depict the 20 most important predictors for our champion RF model in \Cref{fig:varImp}. The rank score is scaled from $0$ to $100$ and is based on the contribution each predictor makes towards the quality of the RF classification model. Not all predictors are equally important for all three maintenance activities. Some play a bigger role in classifying one maintenance activity over the others. It is worth noting that numerous fine-grained source code changes are ranked high in the list, which confirms their contribution to the model's quality.

\begin{table}[H]
\center
\renewcommand{\arraystretch}{1.5}
\caption{The 20 most important features in the best RF compound model, the score is scaled from 0 to 100}
\label{fig:varImp}    
\begin{tabular}{|l|c|c|c|}
\hline
\rowcolor{lightgray} \textbf{Feature} (keyword/fine grained source code change) & \textbf{Addaptive} & \textbf{Corrective} & \textbf{Perfective}   \\ \hline
fix                       &100.00  &100.00  &90.42  \\ \hline
ADDITIONAL\_FUNCTIONALITY   &75.72   &72.07  &75.72  \\ \hline
STATEMENT\_INSERT           &62.17   &40.19  &62.17  \\ \hline
support                    &54.92   &54.92  &53.20  \\ \hline
ADDITIONAL\_OBJECT\_STATE    &42.27   &42.27  &38.01  \\ \hline
add                        &32.00   &32.00  &25.36  \\ \hline
ALTERNATIVE\_PART\_INSERT    &30.04   &16.71  &30.04  \\ \hline
remov                      &27.47   &23.46  &27.47  \\ \hline
DOC\_UPDATE                 &22.51   &26.77  &26.77  \\ \hline
test                       &26.60   &13.62  &26.60  \\ \hline
STATEMENT\_DELETE           &25.98   &25.98  &17.37  \\ \hline
REMOVED\_FUNCTIONALITY      &15.89   &25.51  &25.51  \\ \hline
implement                  &22.97   &22.97  &18.60  \\ \hline
COMMENT\_INSERT             &22.36   &20.48  &22.36  \\ \hline 
PARAMETER\_INSERT           &20.74   &20.74  &18.76  \\ \hline
issu                       &15.04   &18.05  &18.05  \\ \hline
REMOVED\_OBJECT\_STATE       &12.01   &17.98  &17.98  \\ \hline
allow                      &17.11   &17.11  &14.15  \\ \hline
new                        &15.78   &15.78  &10.66  \\ \hline
ADDITIONAL\_CLASS           &15.44   &15.44  &10.82 \\ \hline
\end{tabular}
\end{table}

\section{Applications}
\label{sec:discussion-promise}

Lehman's Laws teach us that a software system will become progressively less satisfying to its users over time, unless it is continually adapted to meet new needs.
The field of software evolution research can be classified into two groups, the first considers the term evolution as a verb while the second as a noun \cite{lehman2000evolution}.
The \textit{verbal} view is concerned with the question of ``how'', and focuses on means, processes, activities, languages, methods and tools required to effectively and reliably evolve and maintain a software system.
The \textit{nounal} view is concerned with the question of ``what'' and investigates the nature of software evolution, as a phenomenon, and focuses on the nature of evolution, its causes, properties, characteristics, consequences, impact, management and control.
Both views are mutually supportive \cite{lehman2000evolution,lehman2003software}. Moreover, they advocate that the verbal view research will benefit from progress made in studying the nounal view, and both are required if the community is to advance in mastering software evolution. We follow this thinking and put forth two applications. 

\subsection{Software Maintenance Activity Explorer}
\label{sec:shinyDemo}

In the spirit of the verbal view \cite{lehman2000evolution} which focuses on studying the means, methods and tools required to effectively evolve a software system, we implement a tool for exploring software maintenance activities aimed to assist practitioners. 
The Software Maintenance Activity Explorer tool \cite{shiny-maintenance-activities} is aimed at providing an intuitive visualization of software maintenance activities over time. We believe this visualization may be useful to project and team managers who seek to recognize inefficiencies and monitor the health of a software project and its corresponding source code repository. The Software Maintenance Activity Explorer was built with Few's \citeyearpar{few2009now} and Cleveland's \citeyearpar{cleveland1985graphical} principles in mind, which advocate for encoding data using visual cues such as variation in size, shape, color, etc'. We chose stacked bar diagrams to visualize data since they allow for an easy comparison both between maintenance activities within a given time frame (e.g., what maintenance activity dominated a given time frame), and between different time frames (e.g., which of the time frames had more maintenance of a given type).
In addition, bar diagrams allow users to quickly detect anomalies such as peaks and deeps in one maintenance activity or another compared to past periods.

\subsubsection*{Project Activity Visualization}

\begin{figure}[H]
    \centering
    \caption{Software Maintenance Activity Explorer's project activity tab}
    \label{fig:demo-project-activity-tab}
    \includegraphics[height=0.85\textwidth,angle=-90]{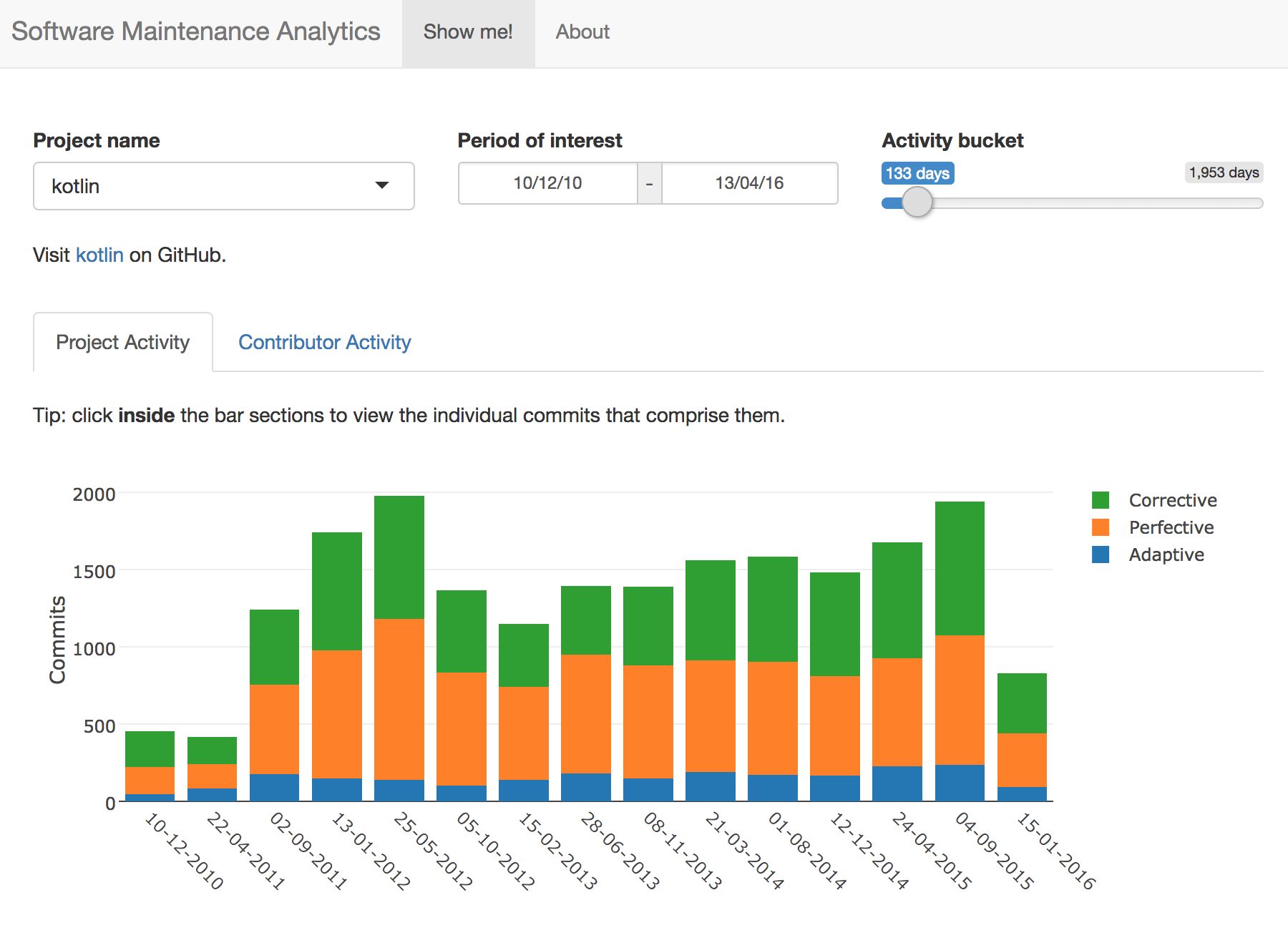}
\end{figure}

The project activity visualization (see \Cref{fig:demo-project-activity-tab}) allows users to examine the volumes of the different maintenance activities over time, and can be sliced and diced according to a specified date range and an activity period (e.g., from date x until date y, in time frames of 28 days). 
The stacked bar plot allows for an easy comparison between the maintenance activity types, as well as trend detection.

\subsubsection*{Developer Activity Visualization}

\begin{figure}[H]
    \caption{Software Maintenance Activity Explorer's developer activity tab}
    \label{fig:demo-developer-activity-tab}
    \includegraphics[height=\textwidth,angle=-90]{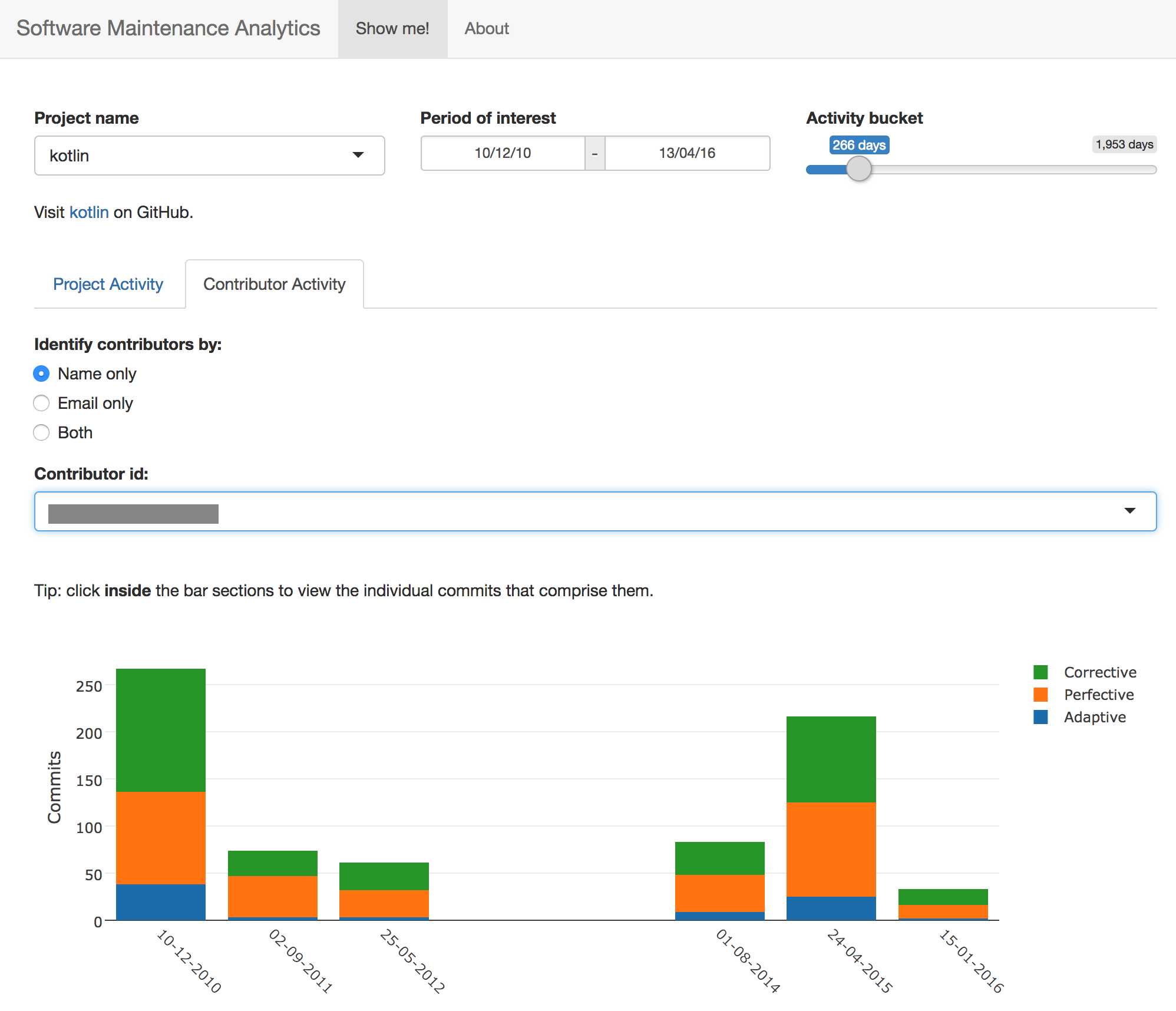}
\end{figure}

The developer activity visualization (see \Cref{fig:demo-developer-activity-tab}) is a segmentation of the data by a specific developer. Users can examine the data for a specific developer, adjusting the period of interest and date range. 
Developers identity can be determined by their name, email or both, a feature that can be useful when developers perform commits using different emails, e.g., when working on an open source project from both their private account and their cooperate account.

\subsubsection*{Publicly Accessible Data}

\begin{figure}[H]
    \centering
    \caption{Software Maintenance Activity Explorer's data exploration tab}
    \label{fig:demo-about-tab}
    \includegraphics[height=0.85\textwidth,angle=-90]{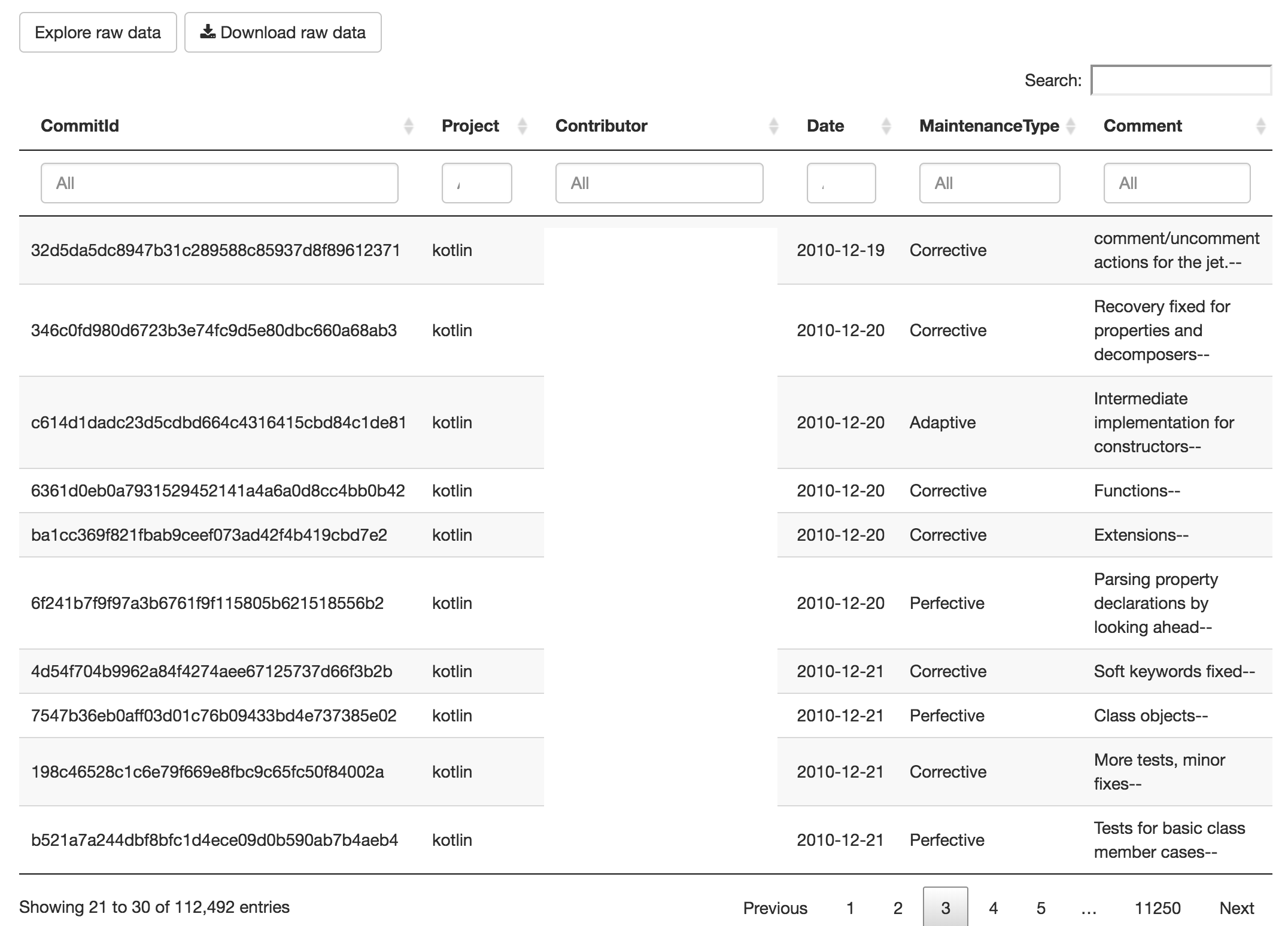}
\end{figure}

The Software Maintenance Activity Explorer's about page provides an option to explore the data in-line (see \Cref{fig:demo-about-tab}), or download it in a CSV format for an offline analysis.

\subsubsection*{Publicly Accessible Code}
The code for this tool is publicly available on GitHub \cite{soft-evo-github}.

We conjecture that a balanced maintenance activity profile, i.e., a profile which includes all three maintenance activity kinds (corrective, perfective, adaptive) may help developers be more effective and engaged with the project they work on. It may also be the case that different project managers will choose different thresholds for what a balanced (or unbalanced) profile is, in the context of their project. Nonetheless, once these thresholds have been set our method provide means to identify opportunities for improvement. This may be of particular interest in open source projects, which tend to heavily rely on community efforts. To that end, well balanced maintenance activity profiles may be something the community needs to drive development forward and ensure that the project gets a fair share of new features, bug fixing, and design improvements - activities which tend to compete for resources in real-world scenarios.

\begin{figure}[H]
  \begin{subfigure}[b]{0.4\textwidth}
    \includegraphics[height=175px]{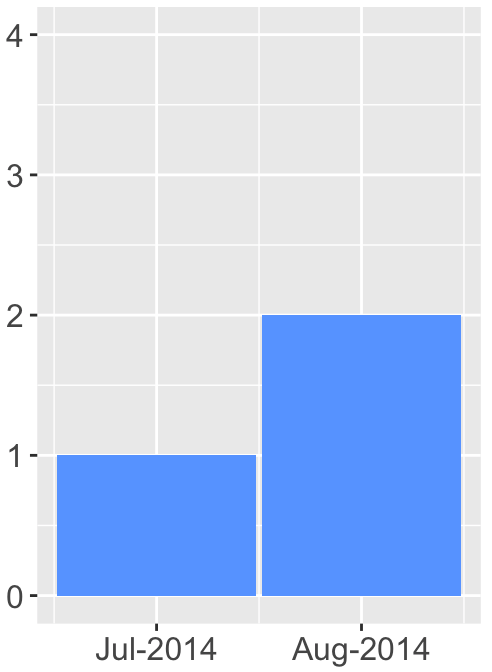}
    \caption{A homogeneous maintenance profile}
    \label{fig:kotlin-only-one}
  \end{subfigure}
  \begin{subfigure}[b]{0.4\textwidth}
    \includegraphics[height=175px]{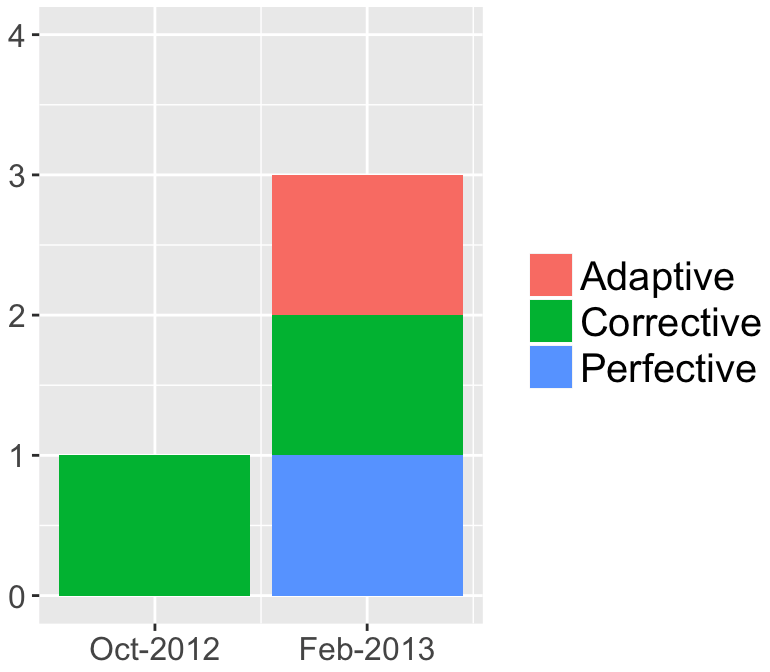}
    \caption{A heterogeneous maintenance profile}
    \label{fig:kotlin-a-c-p}
  \end{subfigure}
  \caption{Maintenance activity profiles for two developers from the Kotlin project}
\end{figure}

We use our dataset and the software maintenance activity explorer to identify homogeneous activity profiles, i.e., profiles of developers who performed only \textit{one} kind of maintenance activity, see \Cref{fig:kotlin-only-one} and \Cref{fig:kotlin-a-c-p}.

The visualization offered by our tool makes it easier to identify these homogeneous maintenance activity profiles and encourage developers to take on a more varied set of tasks.
We performed the homogeneous maintenance activity profiles test for 10 projects (see also \Cref{tbl:prj-info}) in our study and report the results in \Cref{tab:homogeneous-profiles}.
According to our data, the Camel project had an extremely low portion of homogeneous maintenance activity profiles. It may be the case that Camel's contributors were indeed developers who were inclined towards heterogeneous maintenance activities. 
Alternatively, one could suggest a number of possible scenarios. It is possible that the Camel project had a significant number of contributors whose contribution to the project did not include Java code, i.e., it revolved around documentation, configuration files, and so forth. This would mean that the percentage of homogeneous maintenance profiles is actually higher and it might be best to compute it by considering only Java contributors. 
Another possibility is that the number of contributors to the Camel project significantly increased since we had originally processed its commit history. In which case it would be necessary to re-collect and re-process the project's data to produce a more accurate result.

Our analysis indicates that homogeneous maintenance activity profiles were not uncommon in the projects we inspected (see \Cref{tab:homogeneous-profiles}). We believe that unbalanced (i.e., where a significant disproportion between maintenance activities is present), and homogeneous maintenance activity profiles in particular, are an opportunity for managers to reach out to developers and suggest taking on tasks that will balance their maintenance activity profiles.
A possible way to identify suitable tasks would be using projects' task management systems (e.g., a JIRA system) which provide contextual and detailed information about the available tasks.
We also hope that this kind of tool will empower both managers and developers to monitor the ongoing maintenance activities and assist in keeping them varied and balanced. Moreover, such a tool may serve as an alerting mechanism in situations which call for special attention, e.g., when the proportion
of unbalanced maintenance profiles exceeds a given threshold.

\begin{table}[H]
\renewcommand{\arraystretch}{1.3}
\caption[]{Homogeneous maintenance activity profile statistics\footnotemark}
\label{tab:homogeneous-profiles}
\begin{tabular}{|c|c|c|c|c|c|}
\cline{2-6}
\multicolumn{1}{c|}{} & \cellcolor{lightgray} \makecell{Corrective\\ only} & \cellcolor{lightgray} \makecell{Perfective\\only} & \cellcolor{lightgray} \makecell{Adaptive\\only} & \cellcolor{lightgray} \makecell{Homogeneous\\ contributors\\(\% of total, truncated)} & \cellcolor{lightgray} \makecell{Total\\Contributors\footnotemark}\\ \hline
\cellcolor{lightgray} Restlet & 11 & 3 & 2 & 41\% & 39 \\ \hline
\cellcolor{lightgray} Drools & 22 & 20 & 8 & 36\% & 137 \\ \hline
\cellcolor{lightgray} OrientDb & 16 & 14 & 4 & 28\% & 120 \\ \hline
\cellcolor{lightgray} Spring Framework & 25 & 18 & 30 & 25\% & 291 \\ \hline
\cellcolor{lightgray} RxJava & 14 & 24 & 8 & 19\% & 211 \\ \hline
\cellcolor{lightgray} Hbase & 8 & 16 & 8 & 16\% & 189 \\ \hline
\cellcolor{lightgray} Elasticsearch & 79 & 54 & 33 & 14\% & 1103 \\ \hline
\cellcolor{lightgray} Kotlin & 14 & 10 & 6 & 14\% & 356 \\ \hline
\cellcolor{lightgray} Hadoop & 4 & 12 & 1 & 12\% & 137 \\ \hline
\cellcolor{lightgray} Camel & 2 & 1 & 0 & <1\% & 410 \\ \hline
\end{tabular}
\end{table}

\addtocounter{footnote}{-2} 
\stepcounter{footnote}\footnotetext{Due to certain technical difficulties we had to exclude the IntelliJ Community Edition project from homogeneous maintenance activity analysis.}
\stepcounter{footnote}\footnotetext{The total number of contributors is updated as of 2018, maintenance activity profiles were computed as part of the original study conducted in 2016.}

\clearpage

\subsection{Utilizing Software Maintenance Activities to Model Test Counts}
\label{sec:test-modeling}

In the spirit of the nounal view \cite{lehman2000evolution} which investigates the nature of software evolution as a phenomenon, we conduct a study which leverages our method to demonstrate the importance of maintenance activities for modeling the number of tests in a software project (see \Cref{sec:test-modeling}).

Automated testing, and automatic unit tests \cite{hamill2004unit} in particular, is a popular technique for improving software quality. As this technique is gaining popularity and becoming ubiquitous among practitioners it is beneficial to have a good understating of its nature, which as it turns out can be alluding.
\citet{beller2015and} conducted a large-scale field study, where 416 software engineers were closely monitored over the course of five months. Their findings indicate that software developers spend a quarter of their work time engineering tests, whereas they think they test half of their time.

In our previous work \cite{DBLP:conf/icsm/LevinY17} we studied 61 open source projects \cite{test-studied-projects} and established a connection between maintenance activities and test (method and classes) counts in software projects. In this section we extend our previous results and focus on the viability of maintenance activities to modeling the number of test methods and test classes in a software project.

The generalized regression models (GLM, \citet{mccullagh1989generalized, venables2013modern}) we devised were of the following form:

{\large
$${\mathit{Test}^M(\mathit{prj}) = \mathit{Constant^M} + \sum\limits_{i=1}^{|\mathit{Predictors}|}(\mathit{coeff}^M_i * \mathit{predictor}^M_i(\mathit{prj}))}$$ 
}
\noindent where:
\begin{itemize*}[label={}]
    \item ${M \in \{\mathit{Methods}, \mathit{Classes}\}}$ is the test metric we model;
    \item $\mathit{Predictors}$ is the set of predictors;
    \item $\mathit{coeff}^M_i$ are the predictor coefficients;
    \item $\mathit{predictor}^M_i(\mathit{prj})$ are predictor values; and
    \item $\mathit{Constant^M}$ is the model constant.
\end{itemize*}

The corresponding models for $\mathit{Test}^{\mathit{Methods}}$ and $\mathit{Test}^{\mathit{Classes}}$ can be found in \Cref{mod:prjTests}.

All predictors were log transformed to alleviate skewed data, a common practise when dealing with software metrics \cite{shihab2012exploration, camargo2009towards}.
Statistically significant predictors of interest are highlighted in lime-green, and the standard error is reported in parenthesis below the estimated coefficients. 
In addition to the variables we are directly interested in, such as the $\mathit{log(corrective)}$, $\mathit{log(perfective)}$ and $\mathit{log(adaptive)}$ we also use $\mathit{log(LOC)}$, $\mathit{log(age)}$ and $\mathit{log(developers)}$ as control variables, in order to reduce the effect of lurking variables which correlate both with the predictors and the predicted (outcome) variable. Control variables are highlighted in light-bisque.

\begin{table}[bh] 
  \centering 
  \caption[Negative Binomial GLM for test method and test class counts]{Negative Binomial GLM for test method and test class counts \cite{DBLP:conf/icsm/LevinY17}} 
  \label{mod:prjTests} 
\begin{tabular}{@{\extracolsep{5pt}}lcc} 
\\[-1.8ex]\hline 
\hline \\[-1.8ex] 
 & \multicolumn{2}{c}{\textit{Predicted variable:}} \\ 
[+1.2ex]
Predictor & $\mathit{Test}^{\mathit{Methods}}$ & $\mathit{Test}^{\mathit{Classes}}$  \\ 
\hline \\[-1.8ex] 
 \rowcolor{lime} log(corrective) & $-$1.696$^{***}$ & $-$1.351$^{***}$ \\ 
  & (0.314) & (0.285) \\ 
 \rowcolor{lime} log(perfective) & 1.621$^{***}$ & 1.583$^{***}$ \\ 
  & (0.397) & (0.358) \\ 
 log(adaptive) & $-$0.247 & $-$0.173 \\ 
  & (0.366) & (0.329) \\ 
 \rowcolor{Bisque1} log(developers) & 0.318$^{*}$ & 0.105 \\ 
  & (0.182) & (0.163) \\ 
 \rowcolor{Bisque1} log(LOC) & 1.189$^{***}$ & 1.053$^{***}$ \\ 
  & (0.171) & (0.154) \\ 
 \rowcolor{Bisque1} log(age) & 0.770$^{***}$ & 0.686$^{***}$ \\ 
  & (0.205) & (0.185) \\ 
 Constant & $-$12.326$^{***}$ & $-$13.289$^{***}$ \\ 
  & (1.873) & (1.702) \\ 
\hline \\[-1.8ex] 
Number of observations & 61 & 61 \\ 
\hline 
\hline \\[-1.8ex] 
  & \multicolumn{2}{r}{$^{*}$p$<$0.1; $^{**}$p$<$0.05; $^{***}$p$<$0.01} \\ 
\end{tabular} 
\end{table} 

The ANOVA type-\uppercase\expandafter{\romannumeral2\relax} analysis computes the changes in the model given any single predictor is dropped and it therefore does not depend on the order of the predictors in the model. Employing ANOVA type-\uppercase\expandafter{\romannumeral2\relax} analysis helps in avoiding situations where regression models may lead to the conclusion that certain predictors possess greater explanatory powers than others only because they appear first \cite{promise2017-keynote}.
The ANOVA type-\uppercase\expandafter{\romannumeral2\relax} analysis for the predictive models $\mathit{Test^{Methods}}$ and $\mathit{Test^{Classes}}$ can be found in \Cref{tab:anovaTestMethods} \Cref{tab:anovaTestClasses} respectively.
Each row indicates the change in the residual deviance and the ``AIC'' measure (Akaike information criterion, an estimator of the relative quality of statistical models) induced by removing a given predictor from the model. The statistical significance for each row is indicated in the rightmost column.
By inspecting the ``AIC'' column in \Cref{tab:anovaTestMethods} \Cref{tab:anovaTestClasses} we learn which predictors can be \textit{excluded} in order to achieve a lower (better) AIC. By inspecting the ``Deviance'' column we learn a given predictor's contribution to ``explaining'' the predicated variables.
The ``base'' model's deviance and AIC are indicated in the ``none'' row.

It is statistically significant that removing the $\mathit{log(corrective)}$ predictor will result in the model's deviance rising from 72 to 95 and its AIC rising from 993 to 1,015. Higher deviance indicates that the new model will have less explanatory power, and higher AIC indicates that it will be worse than the one it is compared to, i.e., the model where the $\mathit{log(corrective)}$ was present. Similar arguments can be applied to the $\mathit{log(perfective)}$ predictor.
The ANOVA analysis confirms that both perfective and corrective maintenance activities are vital to the model, and an attempt to remove either will significantly and adversely affect the model's quality.

Also worth noting is the LOC predictor, its AIC and deviance indicate that it demonstrates statistically significant high explanatory power in both predictive models. This implies that the size of the project has a considerable effect on the number of test methods and test classes it contains.

\begin{table}[H]
\centering
\renewcommand{\arraystretch}{1.5}
  \caption{ANOVA for $\mathit{Test^{Methods}}$} 
  \label{tab:anovaTestMethods} 
\begin{tabular}{|c|c|c|c|l|c|}
  \cline{2-5}
  \multicolumn{1}{c|}{} & \cellcolor{lightgray} Df. & \cellcolor{lightgray} Deviance & \cellcolor{lightgray} AIC & \cellcolor{lightgray} F value &  \cellcolor{lightgray} Pr($>F$) \\ 
  \hline
\cellcolor{lightgray} \textless none\textgreater  &  & $72.311$ & $993.480$ & $$ & $$ \\ 
\hline
\cellcolor{lightgray} log(corrective) & $1$ & $95.903$ & $1,015.071$ & $17.617$ & \cellcolor{yellow} $0.0001^{***}$ \\ 
\hline
\cellcolor{lightgray} log(perfective) & $1$ & $86.599$ & $1,005.767$ & $10.669$ & \cellcolor{yellow} $0.0018^{**}$ \\ 
\hline
\cellcolor{lightgray} log(adaptive) & $1$ & $72.762$ & $991.930$ & $0.336$ & $0.564$ \\ 
\hline
\cellcolor{lightgray} log(developers) & $1$ & $75.007$ & $994.175$ & $2.013$ & $0.162$ \\ 
\hline
\cellcolor{lightgray} log(loc) & $1$ & $106.675$ & $1,025.843$ & $25.661$ & \cellcolor{Bisque1} $5.077e-06^{***}$ \\ 
\hline
\cellcolor{lightgray} log(age) & $1$ & $84.408$ & $1,003.576$ & $9.033$ & \cellcolor{Bisque1} $0.0040^{**}$ \\ 
\hline
\end{tabular}
\end{table}

\begin{table}[H]
\centering
\renewcommand{\arraystretch}{1.5}
  \caption{ANOVA for $\mathit{Test^{Classes}}$} 
  \label{tab:anovaTestClasses} 
\begin{tabular}{|c|c|c|c|l|c|}
  \cline{2-5}
  \multicolumn{1}{c|}{} & \cellcolor{lightgray} Df. & \cellcolor{lightgray} Deviance & \cellcolor{lightgray} AIC & \cellcolor{lightgray} F value & \cellcolor{lightgray} Pr($>F$) \\ 
  \hline
\cellcolor{lightgray} \textless none\textgreater  &  & $71.873$  & $812.533$ & $$ & $$ \\ 
\hline
\cellcolor{lightgray} log(corrective) & $1$ & $89.012$ & $827.673$ & $12.877$ & \cellcolor{yellow} $0.0007^{***}$ \\ 
\hline
\cellcolor{lightgray} log(perfective) & $1$ & $87.015$ & $825.676$ & $11.377$ & \cellcolor{yellow} $0.0013^{***}$ \\ 
\hline
\cellcolor{lightgray} log(adaptive) & $1$ & $72.125$ & $810.786$ & $0.189$ & $0.665$ \\ 
\hline
\cellcolor{lightgray} log(developers) & $1$ & $72.256$ & $810.916$ & $0.288$ & $0.594$ \\ 
\hline
\cellcolor{lightgray} log(loc) & $1$ & $105.557$ & $844.217$ & $25.308$ & \cellcolor{Bisque1}$5.75e-06^{***}$ \\ 
\hline
\cellcolor{lightgray} log(age) & $1$ & $84.017$ & $822.678$ & $9.124$ & \cellcolor{Bisque1}$0.0038^{**}$ \\ 
\hline
\end{tabular}
\end{table}

\clearpage

\begin{figure}[H]
    \centering
    \caption{Corrective activities and unit tests per 1 LOC for 61 projects (see also \citet{DBLP:conf/icsm/LevinY17})}
    \label{fig:c-vs-test-per-loc}
    \includegraphics[width=\textwidth]{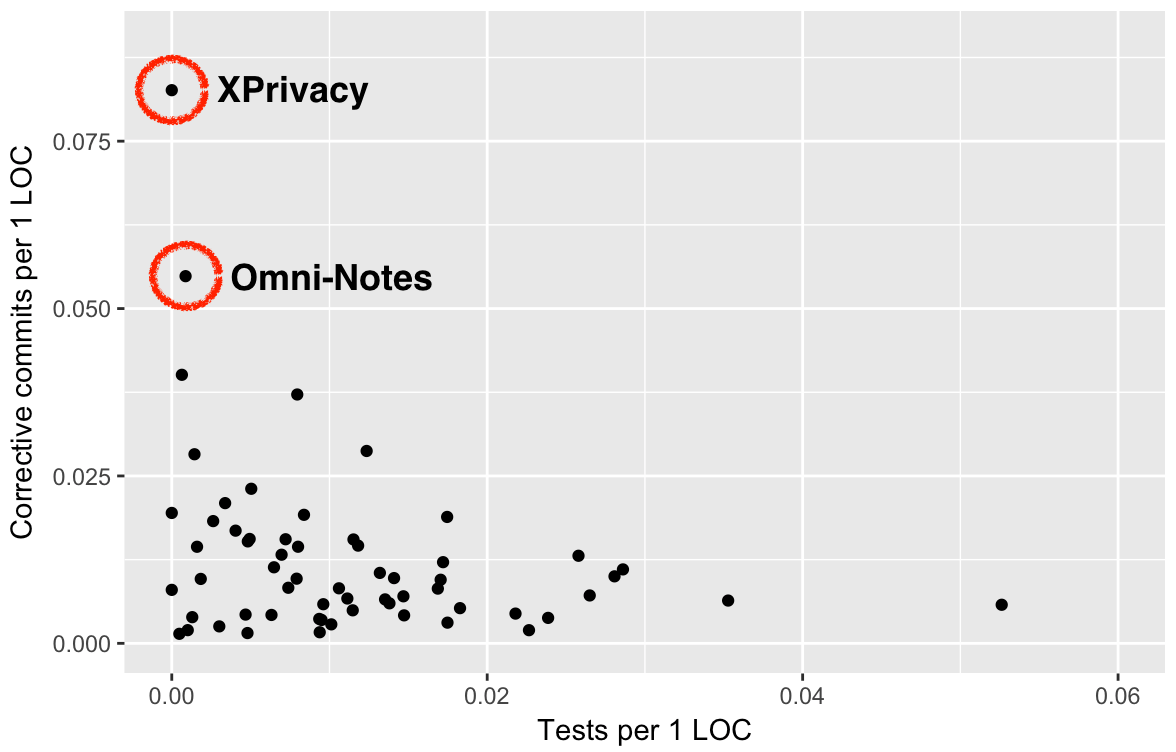}
\end{figure}

Following the insights provided by these test regression models, we performed a deeper inspection of two outlier projects, "XPrivacy" and "Omni-Notes" (see \Cref{fig:c-vs-test-per-loc}), that had extremely high values of corrective activity (per 1 LOC) combined with a low number of tests (per 1 LOC).

Our analysis of XPrivacy did not reveal any unit tests in its codebase. Its README page on GitHub had a designated testing section which revealed that a separate application had been written for testing purposes. The test application's (GitHub) project was nowhere as popular as XPrivacy itself (more than 1.5K stars vs. less than 10 stars) implying it may not have been widely used by developers upon contributing code. 
It is possible that since the test application project was separate from the original application, it was not executed frequently (and automatically) enough, rendering it less effective in preventing defects. This may account for the high amount of corrective activity performed in this project.
Omni-Notes, the second outlier project we inspected, had only 12 tests spread over 8 suites according to our analysis. Its README page on GitHub also had a designated section for testing which specified the build command developers should execute when contributing code.
While the presence of a designated test section in its README page may indicate testing was quite important to the project's owner, the great amount of corrective activity performed in this project may suggest it could have benefited from more unit tests.
Gaining fine grained visibility into anomalies (e.g., as indicated in \Cref{fig:c-vs-test-per-loc}) will allow managers to identify potential issues by examining abnormal values even without knowing the root cause. 
Having identified potential issues, mangers can then shift focus towards investigation and resolution.

To conclude this section, while regression models do not provide means to ascertain causality, the negative correlation between corrective commits and tests (i.e., both methods and classes) is worth considering. Potentially, one could argue that projects with tests may only need little corrective activity due to the high quality of the codebase. The opposite direction, may imply that corrective activity may be required when the test count of a project is low, and the codebase's quality is poor. It is also possible, that test counts and corrective commits do not have a cause and effect relationship at all, in which case they just tend to happen together and are connected via a lurking variable. 
Either of these narratives requires further evidence before it can be reliably established, but to the very least, the empirically evident negative correlation between corrective activity and tests is yet another reminder of the relationship between automated testing and the nature and volume of the maintenance activities a project is likely to require in the future.

\subsection{Future Applications}

\paragraph{Identifying Anomalies In Development Processes} 
The manager of a large software project should aim to control and manage its maintenance activity profiles, i.e., the volume of commits made in each maintenance activity. Monitoring for unexpected spikes in maintenance activity profiles and investigating the reasons (root cause) behind them could assist managers and other stakeholders to plan ahead and identify areas that require additional resource allocation. 
For example, lower corrective profiles could imply that developers are neglecting bug fixing. Higher corrective profiles could imply an excessive bug count. Finding the root cause in cases of significant deviations from predicted values may reveal essential issues the removal of which can improve projects' health. Similarly, exceptionally well performing projects can be a good subject for case studies, so as to identify positive patterns.
    
\paragraph{Improving development team's composition}
Building a successful software team is hardly a trivial task as it involves a delicate balance between technological and human aspects \cite{gorla2004should, guinan1998enabling}. We believe that by using commit classification it would be possible to build reliable developer maintenance activity profiles which could assist in composing balanced teams. We conjecture that composing a team that heavily favors a particular maintenance activity (e.g. adaptive) over the others could lead to an unbalanced development process and adversely affect the team's ability to meet typical requirements such as developing a sustainable number of product features, adhering to quality standards, and minimizing technical debt so as to facilitate future changes.

\section{Threats to validity}
\label{sec:threatsToValidity-promise}

\noindent
\textbf{\textit{Threats to Statistical Conclusion Validity}} are the degree to which conclusions about the relationship among variables based on the data are reasonable.
\begin{itemize}
    \begin{samepage}
        \item
        \underline{Classification Models}. Our commit classification results were based on manually classifying 1151 commits, over 100 commits from each of the studied 11 projects.
        The projects originated from various professional domains such as IDEs, programming languages, distributed database and storage platforms, and integration frameworks. 
        Each compound model was trained using 5-time repeated 10-fold cross validation.
        In addition, our commit classifications evaluations demonstrated $p$-value below 0.01, supporting the statistical validity of the hypothesis accuracy $>$ NIR with high confidence.
    \end{samepage}
    \item
    \underline{Regression Models}. Our dataset for the regression analysis consisted of 61 projects and over 240,000 commits. Both the model coefficients and the predictions were annotated with statistical significance levels to indicate the strength of the signal. Most of the coefficients were statistically significant (${\operatorname{p-value} < 0.01}$).
    To compare distributions we used the Wilcoxon-Mann-Whitney test and reported its high significance level (${\operatorname{p-value} < 0.01}$).\\
    We assume commits are independent, however, it may be the case that commits performed by the same developer share common properties.
\end{itemize}

\noindent
\textbf{\textit{Threats to Construct Validity}} consider the relationship between theory and observation, in case the measured variables do not measure the actual factors.
\begin{itemize}
    \begin{samepage}
        \item
        \underline{Manual Commit Classification}. We took the following measures to mitigate manual classification related errors:
            \begin{enumerate}
                \item Projects' issue tracking systems were used, and often provided additional information pertaining to commits.            
                \item Commits that did not lend themselves to classification due to lack of supporting information were removed from the dataset and replaced by other commits from the same repository (see \Cref{sec:manualLabels}).
                \item A sample of 10\% out of all manually labeled commits was independently classified by both authors. The observed agreement level was 94.5\%, and the asymptotic 95\% confidence interval for the agreement level was [90.3\%, 98.7\%] indicating that both authors agreed about the labels for the vast majority of cases.
            \end{enumerate}
    \end{samepage}
    
    \item
    \underline{Fined-grained Source Code Change Extraction}. ChangeDistiller and the VCS mining platform we have built on top of it are both software programs, and as such, are not immune to bugs which could result in inaccurate or incomplete data.
    
    \item
    \underline{Test Maintenance Classification}. We used a widely practiced conventions and heuristics \cite{mavenTests,zaidman2011studying} for detecting JUnit test methods and test classes. However, the use of heuristics may lead to undetected test maintenance.
    
    \item
    \underline{Data Cleaning}. Prior to devising regression models, we removed extreme data points using a technique suggested in \cite{hubert2008adjusted}. 
    Despite the fact we removed only $\sim$10\% of the data, this process could have introduced bias into the dataset we operated on.
\end{itemize}

\noindent
\textbf{\textit{Threats to External Validity}} consider the generalization of our findings.
\begin{itemize}
\begin{samepage}
        \item
        \underline{Programming Language Bias}. All analyzed commits were in the Java programming language since the tool we used to distill fine grained source code changes (ChangeDistiller) was Java oriented. It is possible that developers who use other programming languages, have different maintenance activity patterns which have not been explored in the scope of this work.
    \end{samepage}
    
    \item
    \underline{Open Source Bias / GitHub}. The repositories studied in this paper were all popular open source projects from GitHub, selected according to the criteria described in \Cref{sec:selectingRepos}. It may be the case that developers' maintenance activity profiles are different in an open source environment when compared to other environments.
    \item \underline{Popularity Bias}. We intentionally selected the popular, data rich repositories. This could limit our results to developers and repositories of high popularity, and potentially skew the perspective on characteristics found only in less popular repositories and their developers. 
    
    \item
    \underline{Limited Information Bias}. The entire dataset, both the training and the test datasets, contained only those commits that we were able to manually classify. At the stage of VCS inspection it can be essentially impossible to actually ascertain the maintenance activities of commits that do not provide enough information traces (comment, ticket id, etc.). The true maintenance activity for such commits may only be known to the developers who made them, and even they may no longer recall it soon after they have moved on to their next task.
    \item \underline{Mixed Commits}. Recent studies \cite{nguyen2013filtering,kirinuki2014hey} report that commits may involve more than one type of maintenance activity, e.g. a commit that both fixes a bug, and adds a new feature. 
    Our classification method does not currently account for such cases, but this is definitely an interesting direction to be  considered for future work (see \Cref{futureWork}).
    
    \item
    \underline{Activity Boundary}. In this work we assume a commit serves as a logical boundary of an activity. It may be the case, that developers perform test maintenance as part of activities that span multiple commits. Such work patterns were not considered in the scope of this work, but are definitely an interesting direction for future work in this area.
\end{itemize}

\section{Summary}
\label{futureWork}

We suggested a novel method for classifying commits into maintenance activities and used it to devise and evaluate a number of models that utilize fine-grained source code changes and the commit message for the purpose of cross-project commit classification into maintenance activities. These models were then evaluated and compared using the accuracy and Kappa metrics with different underlying classification algorithms.
Our champion model showed a promising accuracy of 76\% and Kappa of 63\% when applied on the test dataset which consisted of 172 commits originating from various projects. These results show an improvement of over 20 percentage points, and a relative improvement of over 40\% when compared to previous results (\Cref{currentResults}).
A comparison between the widely used classifier and our champion classifier can be found in \Cref{tab:veryNaiveClassification} and \Cref{tab:bestClassification}, respectively.
Our evaluation was based on studying 11 popular open source projects from various professional domains, from which we manually classified 1151 commits, $\sim$100 from each of the studied projects. The suggested models were trained using repeated cross validation on 85\% of the dataset, and the remaining 15\% of the dataset were used as a test set.

We conclude that the answer to RQ 1. is that fine-grained source code changes can indeed be successfully used to devise high quality models for commit classification into maintenance activities. 

The answer to RQ 2. is that models that utilize source code changes are capable of outperforming the reported accuracy of word frequency based models \cite{hindle2009automatic, amor2006discriminating} from $\sim$60\% to $\sim$75\%, even when classifying cross-project commits.
In addition, we make the following observations based on our study:
\begin{itemize}
    \item Using text cleaning and normalization, our word frequency based models were able to achieve an accuracy of 68-69\% with Kappa of 51-53\% for cross-project commits classification  (see \Cref{tab:model_combos}).
    \item Compound models employing both (commit message) word frequency analysis and source code change types for the task of cross-project commit classification were able to achieve up to 73\% accuracy with Kappa 59\% during the training stage, and up to 76\% accuracy with Kappa of 63\%, considered ''Good`` \cite{altman1990practical}, for the test dataset.
    \item The RF algorithm outperformed the GBM and J48 in classifying cross-project commits (see \Cref{tab:testsetResults} and \Cref{tab:bestClassification}).
\end{itemize}

To explore RQ. 3 we demonstrated two applications for our classification and repository harvesting methods, one in the spirit of the verbal view, and the other in the spirit of the nounal view. 

\begin{itemize}
    \item The Software Maintenance Activity Explorer, a tool that is aimed at providing an intuitive visualization of code maintenance activities over time. It provides users with both project wide, and developer centring views of maintenance activities over various periods of time. We then showed how the software maintenance activity explorer and our dataset can be used to identify homogeneous maintenance activity profiles, which we believe managers should be made aware of and act upon.
    \item  Detecting software projects which may be lacking in tests and potentially require extensive corrective maintenance. The suggested application employs insights obtained from modeling the relationship between commit classification (into maintenance activities) and the number of test methods in a software project.
\end{itemize}

\section{Future Work}

We believe that our methods and results can be leveraged to further explore numerous directions in the field of software evolution and software analytics in particular. 
For example, it would be interesting to learn whether our software maintenance activity explorer could appeal to practitioners working on open source and/or commercial projects. 
It would also be beneficial to learn what real-life tasks they believe this tool can help with, and/or what changes they would like to suggest to make it useful for their needs. 
In addition, it may be of particular interest to get feedback from developers who took part in the projects we analyzed as part of our publicly available version of the software maintenance activity explorer\footnote{Available at \url{https://soft-evo.shinyapps.io/maintenance-activities}.}.

Some commits may involve more than one type of maintenance activity, and some activities may span more than one commit. It would therefore be beneficial to explore whether extended activities and mixed commits lend themselves to automatic and accurate classification.

The availability of an accurate classification model may make it possible to automatically classify an unprecedentedly large number of projects and commit activities. This, in turn, could shed new light on the distribution of maintenance activities in software projects \cite{schach2003determining, lientz1978characteristics}, a subject the research community is yet to agree upon.

\clearpage
\def\UrlFont{\ttfamily\scriptsize}
\def\UrlBreaks{\do\/\do-}
\bibliographystyle{apa}
\bibliography{bibliography}

\end{document}